\documentclass{aastex631}

\usepackage{amsmath}
\usepackage{todonotes}
\usepackage{appendix}

\begin{document}

\title{Solar Neutrino Flux Fluctuations Caused by Solar Gravity Modes}

\correspondingauthor{Yoshiki Hatta}
\email{yoshiki.hatta@isee.nagoya-u.ac.jp}

\author[0000-0003-0747-8835]{Yoshiki Hatta}
\affiliation{Institute for Space-Earth Environmental Research, Nagoya University, \\
Furo-cho, Chikusa-ku, Nagoya, Aichi 464-8601, Japan}
\affiliation{Max-Planck Institute for Solar System Research \\
Justus-von-Liebig-Weg 3, 37077 G\"{o}ttingen, Germany}
\affiliation{National Astronomical Observatory of Japan \\
2-21-1 Osawa, Mitaka, Tokyo 181-8588, Japan}

\author[0000-0003-1572-3888]{Yuuki Nakano}
\affiliation{Faculty of Science, University of Toyama, \\
Gofuku 3190, Toyama, 930-8555, Japan}

\author[0009-0001-9309-2076]{Sho Sugama}
\affiliation{Department of Physics, Faculty of Engineering Science, Yokohama National University, \\
Yokohama 240-8501, Japan}

\author[0000-0002-1932-3358]{Masanobu Kunitomo}
\affiliation{Department of Physics, Kurume University, \\
67 Asahimachi, Kurume, Fukuoka 830-0011, Japan}

\author[0000-0003-1029-5730]{Hiroshi Ito}
\affiliation{Department of Physics, Graduate School of Science, Kobe University\\
Rokkodaicho 1-1, Nada, Kobe, Hyogo 657-8501, Japan}

\author[0000-0001-6583-2594]{Takashi~Sekii}
\affiliation{National Astronomical Observatory of Japan \\
2-21-1 Osawa, Mitaka, Tokyo 181-8588, Japan}
\affiliation{Astronomical Science Program, The Graduate University for Advanced Studies, SOKENDAI, \\ 2-21-1 Osawa, Mitaka, Tokyo 181-8588, Japan}



\begin{abstract}

We evaluate fluctuations in solar neutrino fluxes induced by gravity (g) modes, extending the earlier first-order perturbative framework to second order within the approximation of linear adiabatic oscillation of a spherically symmetric star. We find that the first-order fluctuation is zero due to geometrical cancellation. 
We still find that the second-order fluctuation is non-zero, which consists of time-varying and non-time-varying components. 
The amplitude of the time-varying component is ${\sim} 10^{-9}$ (in relative difference, in the case of $\mathrm{^{8}B}$ neutrino), when we assume the g-mode amplitude parameter $A_{n \ell}$ to be $10^{-5}$, which corresponds to an assumed maximum relative temperature perturbation inside the Sun. 
Thus, individual g-mode detection through neutrino flux measurements lies beyond the sensitivity of present-generation experiments. Notwithstanding this negative result for individual modes, the cumulative second-order effect of a large number of g-modes that originates from the non-time-varying component could be non-negligible. 
In particular, since $A_{n \ell}$ may be related to convection amplitude, which could change in accordance with the solar magnetic activity, the net increase in the neutrino flux, which is proportional to $A_{n \ell}^2$, should also change with the solar activity cycle. Such a long-period variation~(${\sim} 11$~years) in the neutrino flux could thus be interpreted as evidence for a bunch of solar g-modes. Comparison of the theoretical prediction with the solar neutrino measurements by, e.g., Super-Kamiokande, may help us better understand excitation mechanisms of solar g-modes. 

\end{abstract}

\keywords{Helioseismology~(709) --- Solar physics~(1476) --- Solar neutrinos~(1511) --- Nuclear fission~(2323)}


\section{Introduction} \label{sec:intro}

Solar gravity (g) modes have long been sought since the beginning of helioseismology~\citep[e.g.][]{Appourchaux+2010}. 
This is because g-modes have strong sensitivity around the deep radiative region~($r/R_{\odot} < 0.5$) and the g-mode detection enables us to investigate the dynamics and structure around the central region that cannot be robustly inferred only with pressure (p) modes currently measurable. 
There have been multiple claims to detect solar g-modes~\citep[e.g.][]{Severnyi+1976,Brookes+1976,Delache+Scherrer1983,Garcia+2007,Fossat+2017,Fossat+Schmider2018}, but none of them have been widely accepted in the community~\citep[e.g.][]{Appourchaux+2000, Appourchaux+2010, Schunker+2018, Appourchaux+Corbard2019, Scherrer+Gough2019, Boning+2019}. 
A primary difficulty in the g-mode measurements arises from the fact that, as g-modes are evanescent in the convection zone, the g-mode amplitudes decay exponentially in the convection zone and thus the amplitudes at the surface are too small for us to measure via solar surface observations. 
\citet{Gough1985} provided early theoretical estimates of g-mode surface amplitudes and assessed the
prospects for their detection. Recent theoretical studies predict that g-mode amplitudes are smaller than a few $\mathrm{mm} \, \mathrm{s}^{-1}$ in line-of-sight velocity ~\citep[see][and references therein]{Pincon+2021,Belkacem+2022}, requiring us quite a long observational time by optical light~($\sim$ a few decades) in addition to preexisting observations ($\sim$ a few decades) to achieve firm detection of g-modes~\citep{Appourchaux+2010}. 

With the difficulty in the solar g-mode detection via line-of-sight velocity and intensity in mind, other possible observables have been suggested~\citep[e.g.][]{Burston+2008, Polnarev+2009, Lopes+Turck-Chieze2014}. 
One of the promising observables may be neutrino, which is produced by the nuclear reaction in the solar core~\citep{1938PhRv...54..248B, 1939PhRv...55..434B, Kippenhahn+2013,Haxton+2013}. 
The idea is that solar g-modes cause density and temperature fluctuations around the central region, leading to periodic variations in nuclear reaction rates as well as in the resultant neutrino fluxes. 
Possible effects of solar g-mode oscillations on neutrino fluxes were first investigated by~\citet{Gough1991}, showing that g-mode oscillations could produce measurable modulations in the solar neutrino flux. 
Subsequently, \citet{Bahcall+Kumar1993} developed the perturbative framework relating core temperature fluctuations to neutrino production rates. 
These pioneering works derived in the context of the solar neutrino problem to explain the apparent deficit in the observed neutrino flux 
at that time. 
Although the deficit was later explained by neutrino oscillation~\citep{2001PhRvL..86.5651F, 2001PhRvL..87g1301A, 2002PhRvL..89a1301A}, \citet{Gough1991} and \citet{Bahcall+Kumar1993} provided the analytical basis upon which subsequent work has built, leading to numerous concerted studies of helioseismology and solar neutrino, such as calibration of solar models based on neutrino flux measurement and helioseismic observations \citep[e.g.,][]{TC+2001,C+2003,BP+2004,AB+2005,LT13ApJ,Lopes2013Ph}, analysis of a relationship between solar oscillations and neutrino flux fluctuations \citep[e.g.,][]{Lopes2013ApJ,Lopes+Turck-Chieze2014}, and attempts to put constraints on dark matter properties \citep[e.g.,][]{LS10ApJ,LS10Sci,TC+2012}. 
A comprehensive account of how helioseismic and neutrino observations jointly constrain the solar interior has been given by \citet{TCC11}.

In the context of the solar g-mode detection, the relation between the g-mode and solar neutrinos has been 
investigated in great detail by~\citet{Lopes+Turck-Chieze2014}~(hereafter LT14). 
LT14 made the conceptual breakthrough of formulating the neutrino flux fluctuations caused by g-modes based on the assumption of linear adiabatic oscillation. 
By comparing their formulation with the $^8{\mathrm{B}}$ neutrino fluxes observed by the SNO experiment~\citep{2010ApJ...710..540A}, in which no significant periodicity was confirmed, they put an upper limit for g-mode amplitudes, independently from studies using the surface observables such as line-of-sight velocity and intensity. 
LT14\textquoteright{}s result thus highlights a high potential of the solar neutrino as a tool not only to find solar g-modes but also to study, e.g., the excitation mechanism of solar g-modes.

The earlier first-order formalism admits natural extensions, particularly the inclusion of second-order effects arising for nonradial g-modes. 
Actually, as will be shown later in Section \ref{sec:2}, the first-order fluctuation in the \textit{total integrated} neutrino flux vanishes identically for nonradial modes in the adiabatic, spherically symmetric limit, owing to the orthogonality of spherical harmonics. Accordingly, we need to consider the second-order effect to evaluate the net neutrino flux fluctuations.  
In addition, LT14 focused on the SNO data ($^{8}\mathrm{B}$ neutrinos), though we currently have abundant observations of various solar neutrinos, such as $pp$, $pep$, $\mathrm{^{7}Be}$, $\mathrm{^{8}B}$, and CNO-cycle except for $hep$~\citep[see][and references therein]{Bergstrom+2016,ParticleDataGroup:2024cfk}. 

We are also awaiting future neutrino detectors such as Hyper-Kamiokande~\citep{2018arXiv180504163H}, JUNO~\citep{2023JCAP...10..022A}, DUNE~\citep{2015arXiv151206148D, 2019PhRvL.123m1803C}, SNO$+$~\citep{SNO:2021xpa}, and so on. 
In this context, it should be quite valuable for us to evaluate flux fluctuations caused by solar g-modes for neutrinos other than $^{8}\mathrm{B}$. 

The goal of this paper is thus twofold: first, we expand upon the formulation of LT14 taking the second-order effect into account, and second, using the derived expression, we assess the possibility of solar g-mode detection via solar neutrino measurements. 

This paper is structured as follows: in Section~\ref{sec:2}, we present formulations of neutrino flux fluctuations caused by g-modes, assuming linear adiabatic oscillation. 
Using the formulation and state-of-the-art solar models, we numerically evaluate the neutrino flux density fluctuations in Section~\ref{sec:3}. The numerical evaluation is compared with current solar neutrino measurements, based on which we discuss the possibility of solar g-mode detection via solar neutrino measurements in Section~\ref{sec:4}. Finally, we conclude our study and give future prospect in Section~\ref{sec:5}. 

\section{Formulation} \label{sec:2}
In this section, based on the theory of linear adiabatic oscillation of a spherically symmetric star, we derive equations that relate neutrino flux fluctuations to temperature perturbations caused by g-modes. 
We follow the formalism of LT14, extending it to include the angular integration over the full solid angle and the second-order perturbative terms arising from the nonlinear coupling between the temperature eigenfunction and the neutrino production rate. 
After we give a brief outline of the formulation in Section~\ref{sec:2-1}, we derive an explicit expression of the first-order fluctuations in the neutrino flux caused by g-modes and that of the second-order fluctuations in Sections~\ref{sec:2-2} and~\ref{sec:2-3}, respectively. 

Note that expressions derived in this section are applicable to any stars provided that the assumptions of linear adiabatic oscillation, complete neutrino transparency, and the neglect of the time-delay effect are satisfied.
We thus do not necessarily focus on the Sun in this section. 
Specific examples of the solar case will be given in Section~\ref{sec:3}. 
For a comprehensive account of solar g-mode theory and the observational challenges associated with their detection, see also \citet{Appourchaux+2010}.

\subsection{A brief outline} \label{sec:2-1}

A primary goal of this subsection is to present a formal expression that relates the temperature perturbation to the neutrino flux fluctuation. 
The starting point is the assumption of linear adiabatic oscillation~\citep[e.g.][]{Unno+1989,Aerts+2010} to describe g-modes. 
The assumption of linearity ensures that the Eulerian density perturbation~$\rho'$ is related to the displacement vector~$\boldsymbol{\xi}$ of an eigenmode via the linearized mass conservation equation: 

\begin{eqnarray}
\rho' + \nabla \cdot (\rho_0 \boldsymbol{\xi}) = 0, \label{eq1}
\end{eqnarray}

\noindent 
where $\rho_0$ is the density of an equilibrium state and it is a function of the radius alone~(variables with the subscript~$0$ hereafter represent physical quantities of the spherically symmetric equilibrium state). 
Besides, perturbations in the thermodynamic quantities can be related to each other with the adiabatic exponents, e.g.: 

\begin{eqnarray}
\frac{\delta T}{T_0} = 
(\Gamma_{3,0} - 1) 
\frac{\delta \rho}{\rho_0},  \label{eq2}
\end{eqnarray}

\noindent 
where $T_{0}$ and $\Gamma_{3,0}$ are the temperature and the adiabatic exponent. 
We here denote the Lagrangian perturbation 
of a physical quantity~$q$ by $\delta q$. 
The assumption of adiabatic oscillation may be valid deep inside a star because of the relatively longer thermal timescale there 
compared with g-mode periods; for example, the ratio of the local thermal timescale~$\tau_{\mathrm{th}}$ \citep[see Equation (22.2) in][]{Unno+1989} to a typical g-mode period ($\sim $~a few hours to days) is below $10^{-10}$ in the solar core ($r/R_{\odot} < 0.2$).

We then turn to the nuclear reaction rate, which determines the neutrino flux. 
The reaction rate can be approximated as a power-law function of the density and temperature \citep[e.g.][]{Kippenhahn+2013}: 

\begin{eqnarray}
\varepsilon \propto \rho^\beta T^{\eta}. \label{eq3}
\end{eqnarray}

For one-body nuclear reaction~(which is all we consider in this study actually), the power-law index~$\beta$ is~$1$. 

As for the other index~$\eta$, it is generally considered that higher charge nuclei reactions, such as $\mathrm{^{8}B}$ and CNO-cycle, tend to have more temperature dependence corresponding to~$\eta \ge 20$~\citep{1996PhRvD..53.4202B}.

The functional form of~$\varepsilon$ in Equation~(\ref{eq3}) together with Equations~(\ref{eq1}) and~(\ref{eq2}) allows us to write the perturbed nuclear reaction rate~$\delta \varepsilon$ in the following formal expression: 

\begin{eqnarray}
\frac{\delta \varepsilon}{\varepsilon_0} = 
F \biggl ( \frac{\delta T}{T_0} \biggr ), \label{eq4}
\end{eqnarray}

\noindent 
where~$F$ represents a function that relates $\delta \varepsilon$ and $\delta T$ to a certain order, which will be clarified in Sections~\ref{sec:2-2} and~\ref{sec:2-3}. 

Under the assumption that during g-mode oscillations the branching rate of nuclear reactions is unchanged from that in the equilibrium state, the fluctuation in the neutrino flux density $\delta \, \mathrm{ln} \, \phi$ is simply proportional to $\delta \, \mathrm{ln} \, \varepsilon$, i.e., $\delta \, \mathrm{ln} \, \phi = \delta \, \mathrm{ln} \, \varepsilon$, 
resulting in the following expression: 

\begin{eqnarray} 
\frac{\delta \phi}{\phi_0} = 
F \biggl ( \frac{\delta T}{T_0} \biggr ). 
\label{eq5}
\end{eqnarray} 

\noindent 
Thus, the neutrino flux density fluctuation at a point~($\delta \, \mathrm{ln} \, \phi$) is determined by the temperature perturbation caused by g-modes at that point~($\delta \, \mathrm{ln} \, T$). 
In other words, Equation~(\ref{eq5}) is satisfied locally inside a star  
if we assume linear adiabatic oscillation. 
Note that the neutrino flux density~$\phi$ is here defined as the number of neutrinos produced inside a star per unit area, per unit second, and per unit mass, having the units of~$\mathrm{cm}^{-2} \, \mathrm{s}^{-1} \, \mathrm{g}^{-1}$.

What we actually observe can be expressed with the integration of 
the Eulerian perturbation in the neutrino flux density~$\phi'$ throughout the interior, namely: 

\begin{eqnarray}
\Delta \Phi_{\mathrm{g\text{-}mode}} = \int (\phi_0 + \phi') (\rho_0 + \rho') \mathrm{d} V 
- 
\int \phi_0 \rho_0 \mathrm{d} V. \label{eq6}
\end{eqnarray}

\noindent 
When we assume that neutrinos are completely transparent inside a star, which is the case for ordinary main-sequence stars, 
we may consider the whole stellar interior as an integration domain. 
Note that the unit of $\Delta \Phi_{\mathrm{g\text{-}mode}}$ ($\mathrm{cm}^{-2} \, \mathrm{s}^{-1}$) is different from that of $\delta \phi$ ($\mathrm{cm}^{-2} \, \mathrm{s}^{-1} \, \mathrm{g}^{-1}$); the former is the neutrino flux, and the latter is the neutrino flux \textit{density}. 
It should also be noticed that effects of time delay are neglected in the expression above. 
Namely, to derive Equation~(\ref{eq6}), it is assumed that all the neutrinos coming from different regions of a star 
arrive at the observational point at the same time. 
We however would like to emphasize that the time-delay effect is comparable to or smaller than the second-order fluctuations (Appendix~\ref{sec:4-1-1}). 

In Sections~\ref{sec:2-2} and~\ref{sec:2-3}, we specify the function~$F$ and derive explicit expressions of the neutrino flux fluctuations using Equations~(\ref{eq5}) and~(\ref{eq6}). 

\subsection{The first-order fluctuation in the neutrino flux} \label{sec:2-2}

In this section, we derive an expression for the first-order fluctuation in the neutrino flux~$\Delta \Phi_{\mathrm{g\text{-}mode}}$, that is caused by g-mode displacements. 
We mean by first-order that \textquotedblleft{}only the first-order terms in $\delta \phi$ are taken into account\textquotedblright{}, that is: 

\begin{eqnarray}
\frac{\delta \phi}{\phi_0} 
= 
\frac{\delta \varepsilon}{\varepsilon_0} 
= 
\biggl ( \beta (\Gamma_{3,0} - 1)^{-1} + \eta \biggr) 
\frac{\delta T}{T_0} 
= 
c_1 \frac{\delta T}{T_0},  \label{eq7}
\end{eqnarray}

\noindent 
which can be derived from the linearization of Equation~(\ref{eq3}) combined with Equation~(\ref{eq2}). 
We have introduced a constant~$c_{1}$ for simplifying the expression above. 

By inserting Equation~(\ref{eq7}) into Equation~(\ref{eq6}) and retaining only the first-order terms, we have the following lines: 

\begin{eqnarray}
\Delta \Phi_{\mathrm{g\text{-}mode}} (t)
&=& \int_{0}^{R_{\star}} \int_{0}^{\pi} \int_{0}^{2 \pi} 
[ 
\phi'(r,\theta,\psi,t) \phi_0^{-1} 
+ 
\rho'(r,\theta,\psi,t) \rho_0^{-1} 
]
\phi_0
\rho_0 r^2 \mathrm{sin} \theta \mathrm{d}r \mathrm{d}\theta \mathrm{d}\psi 
\nonumber \\ 
&=& \int_{0}^{R_{\star}} \int_{0}^{\pi} \int_{0}^{2 \pi} 
\biggl [ c_1 \delta T(r,\theta,\psi,t) T_{0}^{-1} - \frac{\partial \, \mathrm{ln} \, \phi_0}{\partial r} \xi_{r}(r,\theta,\psi,t) \biggr ] 
\phi_0 \rho_0 r^2 \mathrm{sin} \theta \mathrm{d}r \mathrm{d}\theta \mathrm{d}\psi 
\nonumber \\
\, &\,& \, \, \, \, \, 
+ \int_{0}^{R_{\star}} \int_{0}^{\pi} \int_{0}^{2 \pi} 
\biggl [
\delta \rho(r,\theta,\psi,t) \rho_{0}^{-1} - \frac{\partial \, \mathrm{ln} \, \rho_0}{\partial r} \xi_{r}(r,\theta,\psi,t)  
\biggr ] 
\phi_0 \rho_0 r^2 \mathrm{sin} \theta \mathrm{d}r \mathrm{d}\theta \mathrm{d}\psi, 
\label{eq8}
\end{eqnarray}

\noindent 
where~$r$, $\theta$, $\psi$, $t$, and $R_{\star}$ are the distance from the center, colatitude, azimuthal angle, time, and the stellar radius, respectively. 
The radial component of the displacement vector~$\boldsymbol{\xi}$ is denoted by~$\xi_{r}$. 
In the second line in Equation~(\ref{eq8}), the Eulerian perturbation is converted to the Lagrangian perturbation via the relation, e.g., $\delta \phi = \phi' + (\boldsymbol{\xi} \cdot \nabla) \phi_{0} $, which is correct to first-order. 
Note that physical quantities in the equilibrium state~(those with the subscript $0$) are functions of radius~$r$ alone. 

Let us then specify the functional forms of~$\delta T$ and $\xi_{r}$. 
If we consider a g-mode labeled with $(n,\ell,m)$, where $n$, $\ell$, and $m$ are the radial order, spherical degree, and azimuthal order, respectively, the temperature perturbation~$\delta T_{n \ell m}$ caused by the g-mode can be expressed as: 

\begin{eqnarray}
\delta T_{n \ell m}(r,\theta,\psi,t) = 
\mathrm{Re}[ 
\delta T_{n \ell}(r) 
Y_{\ell}^{m}(\theta, \psi)
e^{-i \omega_{n \ell m} t} ] 
= 
\delta T_{n \ell}(r) 
P_{\ell}^{m}(\mathrm{cos} \theta)
\mathrm{cos}( m \psi - \omega_{n \ell m} t), \label{eq9}
\end{eqnarray}

\noindent 
where the spherical harmonics and associated Legendre polynomials are denoted by~$Y_{\ell}^{m}$ and $P_{\ell}^{m}$. 
The eigenfrequency is represented by~$\omega_{n \ell m}$. 
The corresponding temperature eigenfunction~$\delta T_{n \ell}(r)$ can be obtained once we compute linear adiabatic oscillations of the reference~(spherically symmetric) model. 
Similarly, the radial component of the displacement vector~$\xi_{r}$ is: 

\begin{eqnarray}
\xi_{r,n \ell m}(r,\theta,\psi,t) = 
\xi_{r,n \ell}(r) 
P_{\ell}^{m}(\mathrm{cos} \theta)
\mathrm{cos}( m \psi - \omega_{n \ell m} t). \label{eq10}
\end{eqnarray}

\noindent 
Note that the density perturbation $\delta \rho$ is related to the temperature perturbation~$\delta T$ via Equation~(\ref{eq2}). 
Since we are considering adiabatic oscillation, the eigenfunctions $\delta T_{n \ell}$ and $\xi_{r,n \ell}$ are real, and have the same phase. 
(It should nevertheless be noted that in the nonadibatic regime the eigenfunctions become complex with a depth-dependent phase difference.) 
The explicit functional forms of~$\delta T_{n \ell m}$
and $\xi_{r,n \ell m}$ enable us to rewrite $\Delta \Phi_{\mathrm{g\text{-}mode}}$ as in the following way: 

\begin{eqnarray}
\Delta \Phi_{\mathrm{g\text{-}mode}}(t) 
&=& 
\int_{0}^{R_{\star}} G_{n \ell}(r) \mathrm{d}r 
\int_{0}^{\pi} 
P_{\ell}^{m}(\mathrm{cos} \theta) \mathrm{sin} \theta \mathrm{d}\theta \notag \\
&\times& 
\biggl ( 
\int_{0}^{2 \pi}  
\mathrm{cos} (m \psi) \mathrm{d}\psi \times  \mathrm{cos} (\omega_{n \ell m}t) 
+ 
\int_{0}^{2 \pi} 
\mathrm{sin} (m \psi) \mathrm{d}\psi \times \mathrm{sin} (\omega_{n \ell m}t) \biggr ),
\label{eq11}
\end{eqnarray}

\noindent 
where the function~$G_{n \ell}$ is defined as: 

\begin{eqnarray}
G_{n \ell}(r) = \biggl [ 
[c_1 + (\Gamma_{3,0} - 1)^{-1}] \biggl ( 
\frac{\delta T_{n \ell}}{T_0} \biggr ) 
- \xi_{r,n \ell} \frac{\partial \, \mathrm{ln} \, \phi_0}{\partial r} 
- \xi_{r,n \ell} \frac{\partial \, \mathrm{ln} \, \rho_0}{\partial r} 
\biggr ] 
\phi_0 \rho_0 r^2.
\label{eq12}
\end{eqnarray}

Importantly, Equation~(\ref{eq11}) indicates that the first-order fluctuation in the neutrino flux~($\Delta \Phi_{\mathrm{g\text{-}mode}}$) is zero for $\ell \neq 0$; in the case of $m=0$, the integration over the colatitude is zero due to the orthogonality of the associated Legendre polynomials, and in the case of non-zero~$m$, the azimuthal integration is zero. 
As there is no radial~($\ell = 0$) g-mode in nature, if we focus only on the first-order fluctuation, we cannot observe a net fluctuation in the neutrino flux. 
This is the reason why we would like to expand the formulation to the second-order. 
Note that the first-order result obtained by LT14 corresponds to the local neutrino flux response before angular integration, which remains physically meaningful as a sensitivity measure. We also would like to emphasize that the ``zero-contribution\textquotedblright{} mentioned above is not the case if the assumptions adopted in this section were lifted; stars are aspherical, their oscillations are somewhat nonadiabatic, and the stochasticity may dominate in their mode excitation. 
Although deviations caused by the neglect of these effects are actually estimated to be much smaller than that caused by the time-delay effect, which can be comparable to or smaller than the second-order fluctuations (Appendix~\ref{sec:4-1-1}), we should keep in mind that the ``zero-contribution'' stems from purely mathematical perspectives. 
Regarding this, it is worth noting that the radial weighting function $G_{n \ell}(r)$ peaks in the core, where g-mode amplitudes are maximal. 
Any departure from the idealisations underlying the cancellation, however small, is therefore amplified by the favourable spatial overlap between the g-mode eigenfunction and the neutrino production kernel, as quantified by the form factor $B_{\ell n}$ of LT14. 
Finally, we note that future detectors with improved angular resolution could, in principle, circumvent the angular cancellation entirely, although neither of the future detectors, namely, Hyper-Kamiokande, JUNO, DUNE, and SNO+, has such angular resolution unfortunately.

\subsection{The second-order fluctuation in the neutrino flux} \label{sec:2-3}

As we discuss in the last section, the first-order fluctuation in the neutrino flux that is caused by g-modes is zero (if we consider linear adiabatic oscillation of a spherically symmetric star and neglect the time-delay effect). 
We therefore would like to evaluate the second-order fluctuation in this section. 
In this case, the neutrino flux density fluctuation is: 
\begin{eqnarray}
\frac{\delta \phi}{\phi_0} 
= 
\frac{\delta \varepsilon}{\varepsilon_0} 
= 
c_1 \frac{\delta T}{T_0} 
+ 
\biggl [ 
\frac{\beta (\beta - 1)}{2} 
(\Gamma_{3,0} - 1)^{-2} 
+ 
\beta \eta (\Gamma_{3,0} - 1)^{-1} + 
\frac{\eta (\eta - 1)}{2} 
\biggr ] 
\biggl (\frac{\delta T}{T_0}\biggr )^2 
= 
c_1 \frac{\delta T}{T_0} 
+ 
c_2 \biggl (\frac{\delta T}{T_0}\biggr )^2.   \label{eq13}
\end{eqnarray}
The expression above can be obtained by expanding Equation (\ref{eq3}) around the equilibrium values ($T_0$, $\rho_0$, $\varepsilon_0$, etc.), omitting terms higher than third-order, and then substituting for the relation $\delta \, \mathrm{ln} \, \varepsilon = \delta \, \mathrm{ln} \, \phi$. 
A new constant $c_2$ is introduced for simplifying the notation. 

With Equations (\ref{eq6}) and (\ref{eq13}), we can compute the second-order fluctuation in almost the same manner as we did for evaluating the first-order fluctuation. 
One difference is that 
the eigenfunctions are given as sums of each eigenmode, for instance: 
\begin{eqnarray}
\delta T(r,\theta, \psi,t) 
&=& 
\sum_{n \ell m} \delta T_{n \ell m}(r,\theta, \psi, t) \label{eq16} \\ 
\boldsymbol{\xi}(r,\theta, \psi,t) 
&=& 
\sum_{n \ell m} \boldsymbol{\xi}_{n \ell m}(r,\theta, \psi, t). \label{eq17}
\end{eqnarray}

\noindent We therefore need to be careful about the order of summation and integration in the second-order analysis where couplings among the modes are allowed. 
Note also that, because we are assuming linear adiabatic oscillation, we will persistently use the first-order relation between the Eulerian perturbation and the Lagrangian perturbation, e.g., $\delta \phi = \phi' + (\boldsymbol{\xi} \cdot \nabla) \phi_0$. 

Keeping in mind the points mentioned above, we can rewrite Equation (\ref{eq6}) as follows: 
\begin{eqnarray}
\Delta \Phi_{\mathrm{g\text{-}mode}} = 
\int 
\biggl [ 
\phi_0 + \delta \phi - \xi_r \biggl ( \frac{\partial \phi_{0}}{\partial r} \biggr ) 
\biggr ] 
\biggl [ \rho_0 + \delta \rho 
- \xi_r 
\biggl ( \frac{\partial \rho_{0}}{\partial r}
\biggr )
\biggr ]
\mathrm{d}V 
- \int \phi_0 \rho_0 \mathrm{d}V. \nonumber \\ 
 \label{eq18}
\end{eqnarray}
\noindent Only the perturbed quantities are functions of $(r,\theta,\psi,t)$ and the equilibrium quantities are those of~$r$. 
Substitution of Equations~(\ref{eq2}) and (\ref{eq13}) for Equation~(\ref{eq18}) leads to: 
\begin{eqnarray}
\Delta \Phi_{\mathrm{g\text{-}mode}}
&=& 
\int 
[c_1 (\Gamma_{3,0} - 1)^{-1} + c_2 ] 
\biggl ( \frac{\delta T}{T_0} \biggr )^2 
\phi_0 \rho_0 
\mathrm{d}V \nonumber \\ 
&-& 
\int 
(\Gamma_{3,0} - 1)^{-1} 
\biggl ( \frac{\delta T}{T_0} \biggr )
\xi_r 
\biggl ( 
\frac{\partial \, \mathrm{ln} \, \phi_{0}}{\partial r} 
\biggr )
\phi_0 \rho_0 
\mathrm{d}V 
- 
\int 
c_1
\biggl ( \frac{\delta T}{T_0} \biggr )
\xi_r 
\biggl ( 
\frac{\partial \, \mathrm{ln} \, \rho_{0}}{\partial r} 
\biggr ) 
\phi_0 \rho_0 
\mathrm{d}V \nonumber \\ 
&+& 
\int 
\xi_r^2 
\biggl ( 
\frac{\partial \, \mathrm{ln} \, \phi_{0}}{\partial r} 
\biggr ) 
\biggr ( 
\frac{\partial \, \mathrm{ln} \, \rho_{0}}{\partial r} 
\biggr ) 
\phi_0 \rho_0 
\mathrm{d}V 
\biggr ]
\phi_0 \rho_0 
\mathrm{d}V. 
 \label{eq19}
\end{eqnarray}

By separating variables in a similar way as we did in the last section (see Equations (\ref{eq9}) and (\ref{eq10})), we can specify the functional forms of eigenfunctions that correspond to a certain g-mode labeled with $(n,\ell,m)$ as below: 
\begin{eqnarray}
\delta T_{n \ell m}(r,\theta,\psi,t) &=& 
\delta T_{n \ell}(r) 
P_{\ell}^{m}(\mathrm{cos} \theta)
\mathrm{cos}( m \psi - \omega_{n \ell m} t - \delta_{n \ell m}), \label{eq20}  \\  
\xi_{r,n \ell m}(r,\theta,\psi,t) &=& 
\xi_{r,n \ell}(r) 
P_{\ell}^{m}(\mathrm{cos} \theta)
\mathrm{cos}( m \psi - \omega_{n \ell m} t - \delta_{n \ell m}), \label{eq21}  
\end{eqnarray}

\noindent where~$\delta_{n \ell m}$ has been introduced 
to describe phase differences among modes with different mode indices.  

Finally, introducing Equations~(\ref{eq16}) and~(\ref{eq17}), with the explicit functional forms of the eigenfunctions~(Equations~(\ref{eq20}) and~(\ref{eq21})), into Equation~(\ref{eq19}) allows us to carry out integration of the horizontal part~($\theta$ and $\psi$), ending up with: 
\begin{eqnarray}
\Delta \Phi_{\mathrm{g\text{-}mode}}(t)
&=& 
\sum_{(n, \ell, m \neq 0)} 
\sum_{(n' = n, \ell'=\ell, m' = m)} 
\pi \int_{0}^{R_\star} Q_{(n \ell),(n' \ell')} \mathrm{d}r \nonumber \\
&+&
\sum_{(n, \ell, m \neq 0)} 
\sum_{(n' \neq n, \ell' = \ell, m' = m)} 
\pi \int_{0}^{R_\star} Q_{(n \ell),(n' \ell')} \mathrm{d}r 
\times \mathrm{cos} (o - o') \nonumber \\ 
&+&
\sum_{(n, \ell, m \neq 0)} 
\sum_{(n', \ell' = \ell, m' = -m)} 
\pi (-1)^m \int_{0}^{R_\star} Q_{(n \ell),(n' \ell')} \mathrm{d}r 
\times \mathrm{cos} (o + o') \nonumber \\ 
&+&
\sum_{(n, \ell, m = 0)} 
\sum_{(n', \ell' = \ell, m' = 0)} 
2 \pi \int_{0}^{R_\star} Q_{(n \ell),(n' \ell')} \mathrm{d}r 
\times \mathrm{cos} (o) \mathrm{cos} (o'), \label{eq24}
\end{eqnarray}

\noindent in which $o = o(t) = \omega_{n \ell m}t + \delta_{n \ell m}$ and $o' = o'(t) = \omega_{n' \ell' m'}t + \delta_{n' \ell' m'}$. 
The function~$Q_{(n,\ell),(n',\ell')}$ describes the strength of mode coupling between a mode with~$(n, \ell, m)$ and another mode with $(n', \ell', m')$. 
(We actually see no coupling between modes with different $m$.) 
The definition of $Q_{(n,\ell),(n',\ell')}$ and more details in derivation of Equation~(\ref{eq24}) can be found in Appendix~\ref{App:A}. 

When we look at Equation~(\ref{eq24}), it is clear that the second-order fluctuation in the neutrino flux that is caused by g-modes can be periodic, although there would be numerous periodicities caused by mode coupling, which may render frequency analysis for solar g-mode detection to be cumbersome. 
Another remarkable point in the expression of the second-order fluctuation is that there are non-time-varying components (see, e.g., the first term in the right-hand side of Equation (\ref{eq24})). 
These non-time-varying components might be a key to understanding the neutrino flux variation whose period is comparable to the solar cycle ($\sim 11$ years) \citep[e.g.][]{Bieber+1990}. 
We discuss these points later in Section~\ref{sec:3-3-3}. 

In the next section, using state-of-the-art solar models and the derived expression (\ref{eq24}), we estimate flux fluctuations for various solar neutrinos, e.g., $^8\mathrm{B}$ and $^7\mathrm{Be}$ neutrinos.

\section{Theoretical evaluation of the neutrino flux fluctuation in the case of the Sun} \label{sec:3}
In this section, we focus on the Sun; we utilize the derived expression for the second-order fluctuation (Equation~(\ref{eq24})) to quantify neutrino flux fluctuations caused by solar g-modes.
Because we need the solar eigenfunctions~($\delta T_{n \ell}$ and $\xi_{r,n \ell}$) as well as the solar equilibrium quantities~($\rho_0$, $T_0$, $\phi_0$, etc.) for numerically evaluating Equation~(\ref{eq24}), we firstly demonstrate what spherically symmetric equilibrium models of the Sun are used in Section~\ref{sec:3-1} and how we calculate linear adiabatic oscillation of the models in Section~\ref{sec:3-2}. 
We then show expected fluctuations in neutrino fluxes 
in Section~\ref{sec:3-3}. 

Note that, because our main goal in this study is to compare theoretical evaluations with the current observations, we here focus on $\mathrm{^{8}B}$ and $\mathrm{^{7}Be}$ neutrinos, only with which reasonable comparisons are feasible as will be shown in Section~\ref{sec:4}. 
We nevertheless give the results of theoretical evaluations for the CNO neutrinos, i.e., $^{13}\mathrm{N}$, $^{15}\mathrm{O}$, and $^{17}\mathrm{F}$ in Appendix~\ref{App:CNO}.

\subsection{Solar models} \label{sec:3-1}

The equilibrium quantities~($\rho_0$, $T_0$, and $\phi_0$) can be obtained once we construct a spherically symmetric model of the Sun~\citep[e.g.][]{JCD+1996,Viny+2017}. 
Since solar models are still actively discussed in the community~\citep[see][and references therein]{Orebi-Gann+21, Buldgen+2025c} we use the following four models from~\citet[][see their table A.1 for details]{Kunitomo+22}: SSM-GS98, SSM-A09, K2-A2-12, and K2-MZvar-A2-12. 
The first two models are the solar standard models~(SSMs) with the old, high-$Z$ abundances~\citep{GS98} and the low-$Z$ ones~\citep{Asplund+09}, respectively, where $Z$ is metallicity.
In the latter two models, we consider the low-$Z$ abundances, protosolar accretion phase, and opacity increase around the base of the surface convective zone, which significantly improves the fit to helioseismic sound speed constraints~\citep[see, e.g.,][]{
Kunitomo+Guillot21, Buldgen+25b}.
In the model K2-A2-12, the composition of the accreted gas is constant with time, while a variable composition of accretion is assumed in K2-MZvar-A2-12 due to planet formation processes~\citep{Kunitomo+Guillot21}. The accretion composition can increase the core metallicity and thus can significantly affect neutrino fluxes. Our fiducial model K2-MZvar-A2-12 reproduces the surface metallicity, sound speed profile, and neutrino fluxes well simultaneously~\citep{Kunitomo+22}.

\begin{figure}[t]
\begin{center}
\includegraphics[scale=0.42]{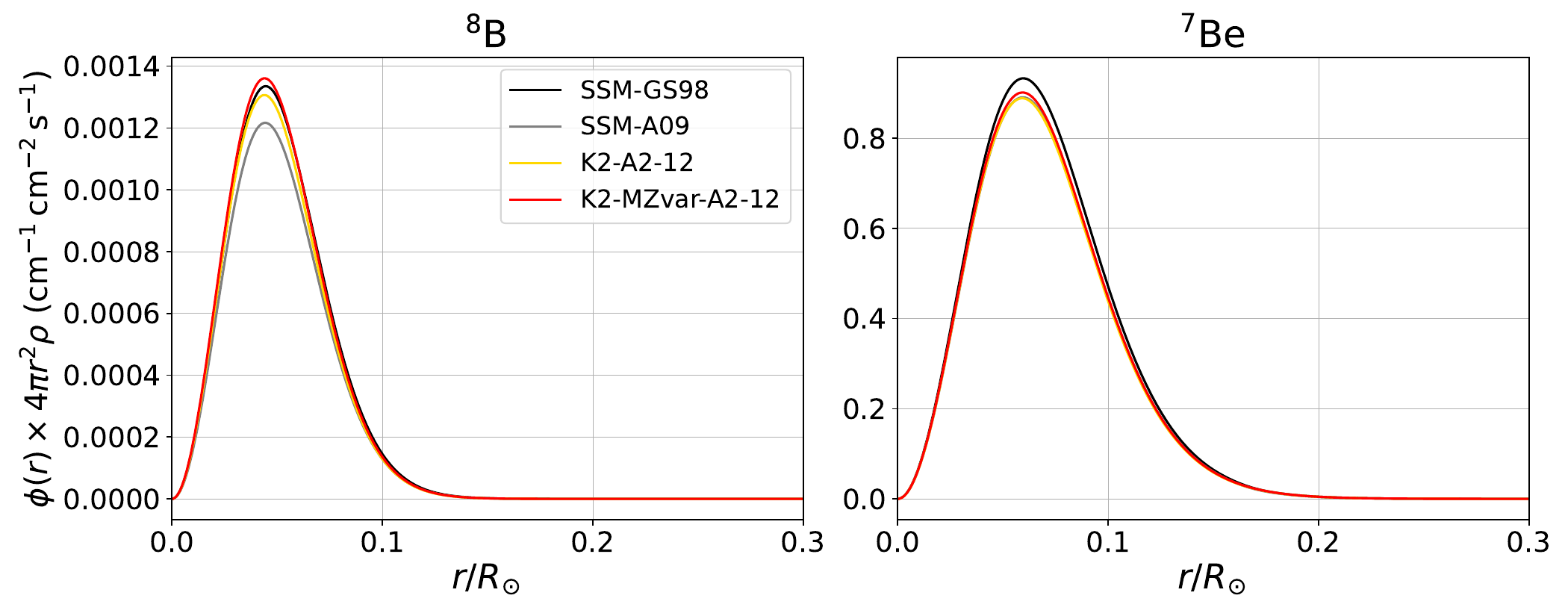}
\caption{\footnotesize Flux per unit length, i.e., $\phi(r) \times 4 \pi r^2 \rho$, 
of $^{8}\mathrm{B}$~(left) and $^{7}\mathrm{Be}$~(right) neutrinos as a function of the fractional radius. 
Different colors represent the four solar models used in this study, namely, SSM-GS98 in black, SSM-A09 in grey, K2-A2-12 in yellow, and K2-MZvar-A2-12 in red. 
(See the main text for more information on the models.) 
A primary difference among the models is in the abundance of heavier elements. }
\label{fig:1}
\end{center}
\end{figure}

To illustrate differences among the four models described above, we show in Figure~\ref{fig:1} internal profiles of the neutrino flux per unit length~(i.e. $\phi(r) \times 4 \pi r^2 \rho$) around the core region of the models. (A similar figure for the CNO neutrinos is given in Appendix~\ref{App:CNO}.) 
We see little differences in where peaks of the neutrino flux densities are located, which reflects the fact that these models have similar temperature profiles in the core region; the ``structural differences" are of the order of a few $\%$~\citep{Kunitomo+22}. 
On the other hand, values of the flux density peaks are different among the models. The difference is attributed to that in the metallicity around the core region as discussed in the last paragraph. 

\begin{figure}[t]
\begin{center}
\includegraphics[scale=0.42]{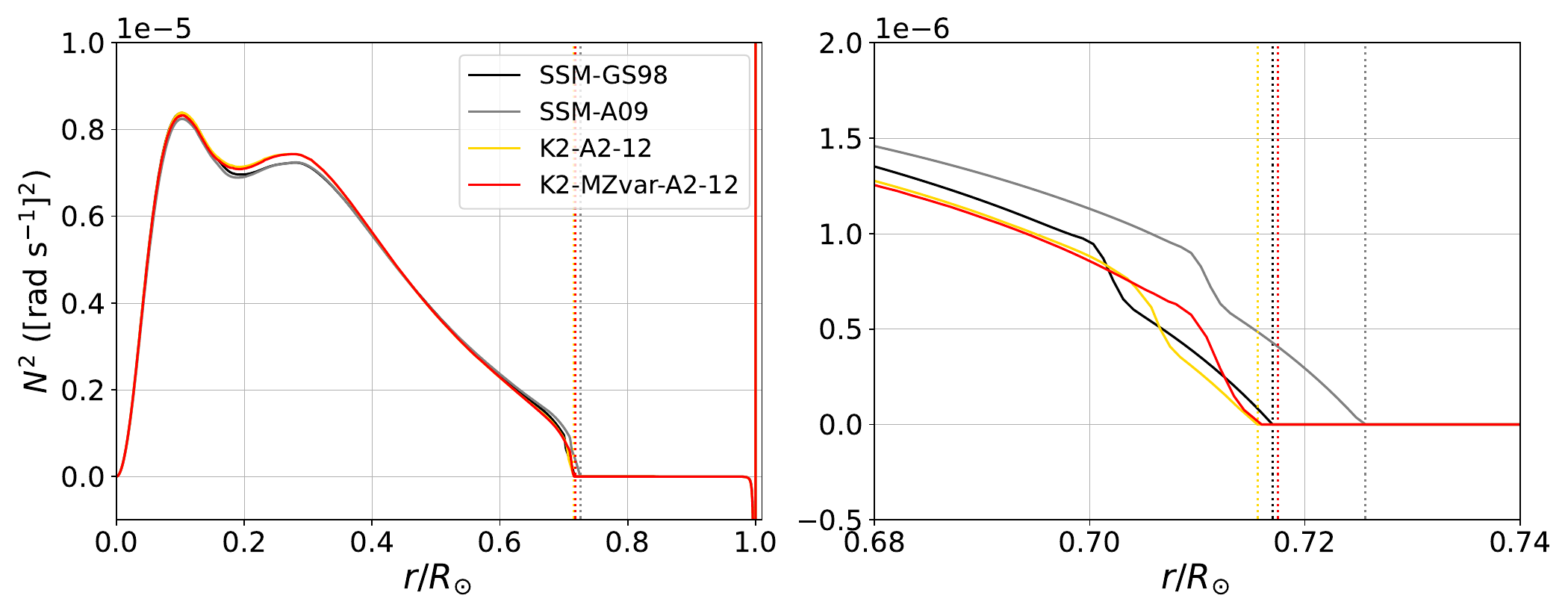}
\caption{\footnotesize Squared Brunt-V\"{a}is\"{a}l\"{a} frequencies $N^2$ (in units of $[\mathrm{rad} \, \,  \mathrm{s}^{-1}]^2$) as a function of the fractional radius. 
The radiative (convective) region is defined as a region where $N^2$ is positive (negative). 
(Note that $N^2$ is almost zero in most of the convective envelope.) 
Different colors represent the four solar models in the same way as in Figure \ref{fig:1}. 
The left panel shows the whole stellar interior, while the right panel shows the close-in look around the bottom of the convective envelope. 
The locations of the bottom of the convective envelope are indicated by dotted lines, in which we see slight differences among the four solar models. }
\label{fig:N2}
\end{center}
\end{figure}

As another example to highlight the differences among the four models, we also show in Figure \ref{fig:N2} the Brunt-V\"{a}is\"{a}l\"{a} frequency (BV frequency hereafter), which is defined as: 
\begin{eqnarray}
N^2 = -g \biggl ( \frac{\mathrm{d} \, \mathrm{ln} \, \rho}{\mathrm{d}r} - \frac{1}{\Gamma_1} \frac{\mathrm{d} \, \mathrm{ln} \, p}{\mathrm{d}r} \biggr ), \label{eqN2}
\end{eqnarray}
where $g$, $p$ and $\Gamma_1$ are the gravitational acceleration, the pressure and the first adiabatic exponent, respectively. 
The BV frequency indicates the convective stability, and it is thus closely related to the g-mode properties \citep[e.g.][]{Unno+1989}. 
Again, we see only slight differences; $N^2$ differs at most a few $\%$, and the locations of the bottom of the convective envelope (the dotted lines) differs at most $1\%$. 
It should nevertheless be noted that these slight differences lead to differences in g-mode eigenfunctions \citep[e.g.,][]{Berthomieu+1990}, consequently affecting the neutrino flux fluctuations. 
We will discuss this point later in the following sections.  

Note that all the four solar models used in this study have not included the effects of rotation whose roles in solar evolution are actively debated \citep{Buldgen+2025c}. 
We have therefore used a solar model where effects of rotationally induced mixing are included \citep{Kunitomo+2025} to numerically evaluate neutrino flux fluctuations (as is explained later in the following sections), but we see almost no differences in the evaluation, which is mainly due to the fact that the structural differences caused by the rotational effects are within the spread we see in the other four solar models. 
Thus, we will hereafter not discuss the solar model with rotational effects taken into account. 

\subsection{Linear adiabatic oscillations of the solar models} \label{sec:3-2}
For each solar model, we have calculated linear adiabatic oscillation via GYRE~\citep[version 6.0.1,][]{Townsend+Teitler2013} to obtain the eigenfunctions~($\xi_{r,n \ell}$ and $\delta T_{n \ell}$). 
We have adopted default settings in GYRE for boundary conditions, where the inner boundary conditions are given by the regularity-enforcing conditions, and the outer boundary conditions are described by the vanishing surface density. 
The ranges of radial orders and spherical degrees of the modes thus computed are $-8 \le  n \le -1$ and $\ell=1,\, 2,\, 3$, respectively, which correspond to one hour to several hours in period. 
When we focus on a mode, the period differences seen among the different models are small~($\sim$ a few $\%$), which are compatible with the structural differences seen in the models. 
Note that effects of rotation, magnetic fields, and asphericity have not been taken into account in the oscillation computations, although the Sun is characterized by such aspherical properties (rotation, the aspherical BV frequency $N(r,\theta)$ that is related to differential rotation, possible magnetic fields around the tachocline, etc.), which could modify the g-mode eigenfunctions. 
Assessing these aspherical effects on the oscillation computations is beyond the scope of this paper, and will be discussed in the future work. 

We then would like to remark on how we determine amplitudes of the eigenfunctions; because the basis of our analysis is the assumption of linear oscillation, we have to somehow determine the amplitudes for numerical evaluation of Equation~(\ref{eq24}). 
We first normalize~$\delta T_{n \ell}$, which is obtained as a result of computations by GYRE, by a constant~$\delta T_{n \ell, 0} > 0$ so that the maximum of $|\delta T_{n \ell}| / \delta T_{n \ell, 0}$ is unity. 
The other eigenfunction~($\xi_{r,n \ell}$) are normalized with $\delta T_{n \ell, 0}$ as well. 
Then, we introduce an amplitude parameter $A_{n \ell}$ and multiply the normalized eigenfunctions by $A_{n \ell}$ to quantify the amplitudes. 
Physically speaking, $A_{n \ell}$ corresponds to the assumed maximum relative temperature perturbation inside the Sun. 
Specific values of $A_{n \ell}$ will be given in the following section. 

\begin{figure}[t]
\begin{center}
\includegraphics[scale=0.40]{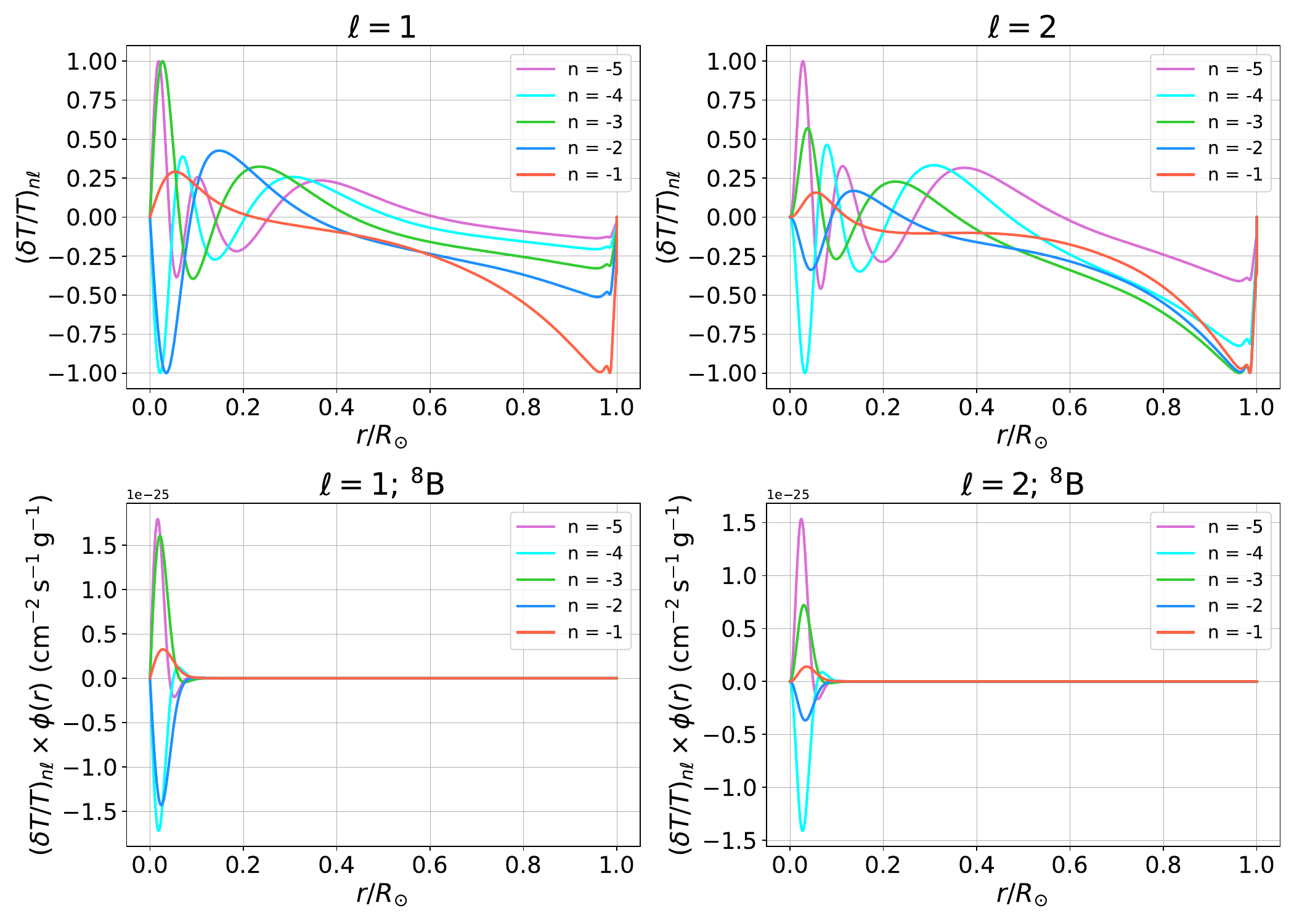}
\caption{\footnotesize Eigenfunctions of the temperature perturbations~$(\delta \,  \mathrm{ln} \, T)_{n \ell}$ (the top row), and $(\delta \,  \mathrm{ln} \, T)_{n \ell} \times \phi$ for $^{8}\mathrm{B}$ neutrino (the bottom row), which is analogous to the gravito-neutrino eigenfunction introduced by LT14, as a function of the fractional radius. 
The left and right panels show dipole~($\ell = 1$) and quadrupole~($\ell = 2$) modes, respectively, 
where radial orders are indicated by different colors. 
These eigenfunctions are obtained by solving linear adiabatic oscillation of the fiducial model, i.e. K2-MZvar-A2-12. 
The maximum amplitude of $(\delta \,  \mathrm{ln} \, T)_{n \ell}$ is normalized to be unity. 
More details on, e.g., the way of normalization, can be found in the main text. 
}
\label{fig:2}
\end{center}
\end{figure}

The top row of Figure~\ref{fig:2} shows examples of the normalized temperature perturbations $\delta T_{n \ell}$ for dipole~($\ell=1$, left panel) and quadrupole~($\ell = 2$, right panel) modes in the case of our fiducial model~(K2-MZvar-A2-12).  
It is seen that, for dipole~(quadrupole) modes with the radial order $n = -2$ to $-5$ ($n = -4$ and $-5$), highest peaks of $\delta T_{n \ell}$ are located around the central region~($r / R_\odot < 0.1$), which corresponds to where the neutrino flux densities are the largest inside the Sun (see Figure~\ref{fig:1}), possibly leading to the net neutrino flux fluctuations as indicated by Equation~(\ref{eq24}). 
This can qualitatively be confirmed by having a look at a product between the temperature perturbation ($\delta T_{n \ell}$) and the equilibrium neutrino flux ($\phi$), which is somewhat analogous to the gravito-neutrino eigenfunction introduced by LT14 (the bottom row in Figure \ref{fig:2}). 
It is seen that relatively higher-order g-modes have stronger sensitivity around the core region. 

As for the other modes with lower radial orders, although their temperature eigenfunctions have amplitudes in the core region as well, the highest peaks are located in the near-surface region (see, for example, the orange curve in the top-left panel of Figure~\ref{fig:2}). 
The significant amplitudes of the relatively low-order modes in the envelope are due to the mixed-mode nature of these modes; their frequencies~(of about $100 \, \mu \mathrm{Hz}$) are high enough for them to behave as p-modes in the envelope \citep[see, e.g.,][]{Pincon+2022}. 
Consequently, the peak values of $\delta T_{n \ell} \times \phi$ are smaller for these relatively low-order modes (see, for example, the orange curve in the bottom-left panel of Figure~\ref{fig:2}). 
It is generally the case that g-mode amplitudes are more concentrated around the core region as the radial order~(the mode period) becomes larger~(longer). 
As will be discussed in the following section, the eigenfunction profiles determine the amount of neutrino flux fluctuations. 
Note that there are only small differences in the normalized eigenfunctions of the four models, which can be explained by small structural differences among the models~($\sim$ a few $\%$) as discussed in Section~\ref{sec:3-1}. 

\subsection{Expected fluctuations in the neutrino flux caused by solar g-modes} \label{sec:3-3} 
Using the eigenfunctions~($\delta T_{n \ell}$ and $\xi_{r,n \ell}$), amplitude parameter~$A_{n \ell}$, and equilibrium quantities~($\rho_0$, $T_0$, and $\phi_0$) described in the last two sections, we numerically evaluate neutrino flux fluctuations (Equation~(\ref{eq24})). We start with a simple case where we evaluate the neutrino flux fluctuation caused by a single g-mode with $m=0$ in Section~\ref{sec:3-3-1}, after which we present a more complex case where couplings among a few dozen modes are considered in Section~\ref{sec:3-3-2}.

\subsubsection{Simple case: neutrino flux fluctuations caused by a single g-mode} \label{sec:3-3-1} 
In order to clearly see relations between the perturbations caused by g-modes and the neutrino flux fluctuations, we begin with the simplest case, where the neutrino flux fluctuation is caused by a single g-mode with $(n, \ell, m=0)$. 
In this case, the fluctuation $\Delta \Phi_{\mathrm{g\text{-}mode}}$ is expressed as: 
\begin{eqnarray}
\Delta \Phi_{\mathrm{g\text{-}mode}}
&=&
2 \pi \int_{0}^{R_\odot} Q_{(n \ell),(n \ell)} \mathrm{d}r 
\times \mathrm{cos}^2 (o) 
= 
\pi \int_{0}^{R_\odot} Q_{(n \ell),(n \ell)} \mathrm{d}r 
+ 
\pi \int_{0}^{R_\odot} Q_{(n \ell),(n \ell)} \mathrm{d}r 
\times \mathrm{cos}(2o). \label{eq26}
\end{eqnarray}
The definitions of the variables are given in the last subsection. 
As it is seen in the rightmost expression in Equation~(\ref{eq26}), the fluctuation consists of a non-time-varying component and time-varying one (with the angular frequency~$2 \omega_{n \ell 0}$). 
Physically speaking, the non-time-varying (DC) part represents the mean neutrino flux increase due to the quadratic sensitivity of $\varepsilon$ to temperature, whilst the oscillatory part modulates at twice the mode frequency.

\begin{figure}[t]
\begin{center}
\includegraphics[scale=0.30]{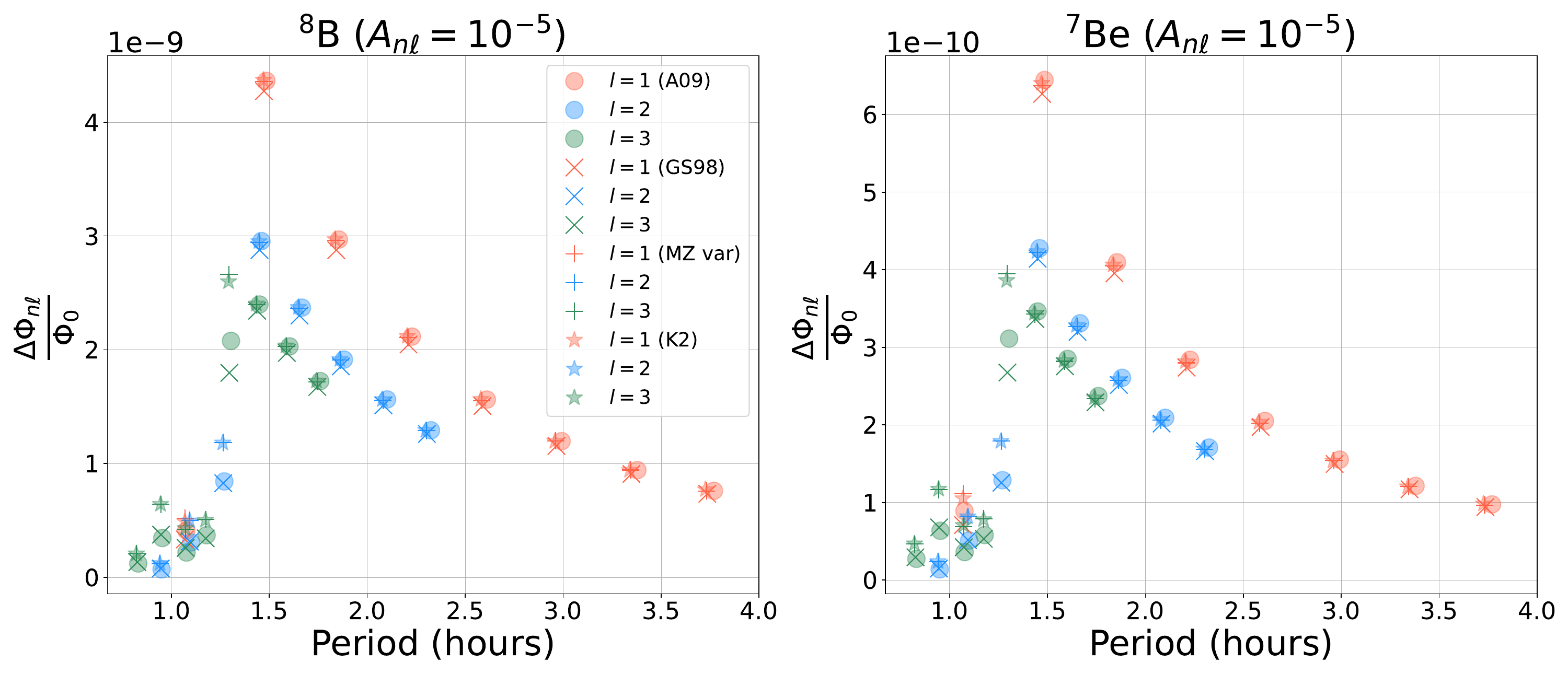}
\caption{\footnotesize Relative fluctuation in the neutrino flux $\Delta \Phi_{n \ell} / \Phi_0$, that is caused by a single g-mode labeled with $(n,\ell)$, as a function of the g-mode period (see, for instance, the first term in the right-hand side of Equation (\ref{eq26})), 
for $^8{\mathrm{B}}$ (left) and 
$^7{\mathrm{Be}}$ (right) neutrinos. 
The numerical evaluations are carried out based on the four solar models described in Section \ref{sec:3-1}, 
which are represented by circles (SSM-A09), crosses (SSM-GS98), plus signs (K2-MZvar-A2-12), and stars (K2-A2-12). 
Results of dipole, quadrupole, and octupole modes are colored in red, blue, and green, respectively. 
The value of the g-mode amplitude parameter $A_{n \ell}$ is assumed to be $10^{-5}$. 
}
\label{fig:3}
\end{center}
\end{figure}

Figure~\ref{fig:3} shows relative amplitudes of neutrino flux fluctuations, namely, the integration~$\pi \int Q_{(n \ell),(n \ell)} \mathrm{d}r$ divided by the total equilibrium neutrino flux~$\Phi_0 = 4 \pi \int \phi_0 \rho_0 r^2 \mathrm{d} r$, as a function of g-mode periods, that are computed for the different four solar models~($^8\mathrm{B}$ and $^7\mathrm{Be}$ neutrinos shown in the left and right panels, respectively). 
The radial order and spherical degree range from $-8$ to $-1$ and from $1$ to $3$, respectively. 
The value of the amplitude parameter $A_{n \ell}$ is assumed to be $10^{-5}$. 
Note that such a uniform distribution of $A_{n \ell}$ across all g-modes is rather a simplified assumption~\citep[see LT14;][for more discussions on possible amplitudes of the temperature perturbation caused by g-modes]{Bahcall+Kumar1993,Belkacem+2009,Pincon+2021,Belkacem+2022}.

As for the power-law indices $\beta$ and $\eta$ (see Equation (\ref{eq3})), we have assumed that $\beta=1$ (because we are considering one-body reactions in this study), and that $\eta = 24$ and $11$ for $^8$B and $^7$Be neutrinos, respectively~\citep{1996PhRvD..53.4202B}.

One prominent feature in the evaluation result (Figure \ref{fig:3}) is that the relative amplitudes $\Delta \Phi_{\mathrm{g\text{-}mode}} / \Phi_0$ are of the order of $10^{-10}-10^{-9}$; apparently, the second-order fluctuations thus evaluated are too small for any existing neutrino detectors to measure. 
This is actually expected because what we have evaluated are the second-order fluctuations rather than the first-order ones as those evaluated by LT14. 
Note that, once we evaluate the integration in Equation~(\ref{eq24}), we can readily evaluate the neutrino flux fluctuations for different $A_{n \ell}$. 
This is because $\Delta \Phi_{\mathrm{g\text{-}mode}} / \Phi_0$ is proportional to $(A_{n \ell})^2$ when we assume $A_{n \ell}$ is constant. Another feature is a tendency that 
the relative amplitude $\Delta \Phi_{\mathrm{g\text{-}mode}} / \Phi_0$ is smaller for shorter periods ($< 1.5$ hours).
This period dependence can be explained by relatively small amplitudes of the eigenfunctions for the short-period modes as we discussed in Section~\ref{sec:3-2} (see also, e.g., the orange curve in the left panel in Figure~\ref{fig:2}). 
In other words, these modes are less sensitive to the region where the neutrino flux density $\phi_0$ is large (see Figure~\ref{fig:1}). 
Such mixed-mode nature can also explain some spreads in Figure \ref{fig:4}; namely, numerical evaluations for $\ell = 2$ (blue) and $\ell = 3$ (green) modes significantly differ among the models in a period range from $1.2$ hours to $1.4$ hours. 
This may in particular originate from a difference in the BV frequency profiles and that in the locations of the bottom of the convective envelope (see Figure \ref{fig:N2}); the depth of a tunneling region differs, and thus, the coupling strength between g- and p-modes is also different among the models, modifying the mixed-mode eigenfunctions. 
Still, the relative differences in the numerically evaluated neutrino flux fluctuations are at most $\sim 20-30 \%$, and not so drastic.  
We also would like to mention that there is almost no model dependence in the numerical evaluations for high-order modes with periods longer than $1.5$ hours. 
This is because these high-order modes are established well inside the g-mode cavity, and not strongly affected by differences in the locations of the bottom of the convective envelope. 

\subsubsection{More complex case: neutrino flux fluctuations caused by dozens of g-modes} \label{sec:3-3-2} 
Although discussions in the last subsection are helpful for articulating the relation between a g-mode oscillation and the resultant neutrino flux fluctuation, there should in reality be numerous g-modes in the solar core, causing couplings among them. 
In this subsection, we numerically evaluate Equation~(\ref{eq24}) taking the mode couplings into account. 
We use the eigenfunctions computed in Section \ref{sec:3-2}, namely, those labeled with $n$ and $\ell$ ranging from $-1$ to $-8$ and $1$ to $3$, respectively. 
Following the last subsection, we assume the amplitude parameter~$A_{n \ell}$ to be $10^{-5}$, and the same values are used for the power law indices~$\beta$ and $\eta$. 
We also consider modes with different azimuthal order~$m$ in a range $- \ell \le m \le \ell$, and thus, the total number of g-modes considered is around~$100$. 
The phase differences $\delta_{n \ell m}$ are sampled from the uniform distribution in a domain between $0$ and $2 \pi$. 
To compute the eigenfrequency $\omega_{n \ell m}$, we only consider the first-order rotational splitting under the asymptotic assumption, i.e., $\omega_{n \ell m} = \omega_{n \ell 0} + m[1 - \ell (\ell +1)^{-1}] \Omega_{\mathrm{c}}$ \citep[see, e.g.,][]{Appourchaux+2010}, where $\omega_{n \ell 0}$ is given as a result of the computation by GYRE, and $\Omega_{\mathrm{c}}$ is the rotation rate of the solar core, which is assumed to be $\Omega_{\mathrm{c}} / 2 \pi =440$ nHz. 

\begin{figure}[t]
\begin{center}
\includegraphics[scale=0.28]{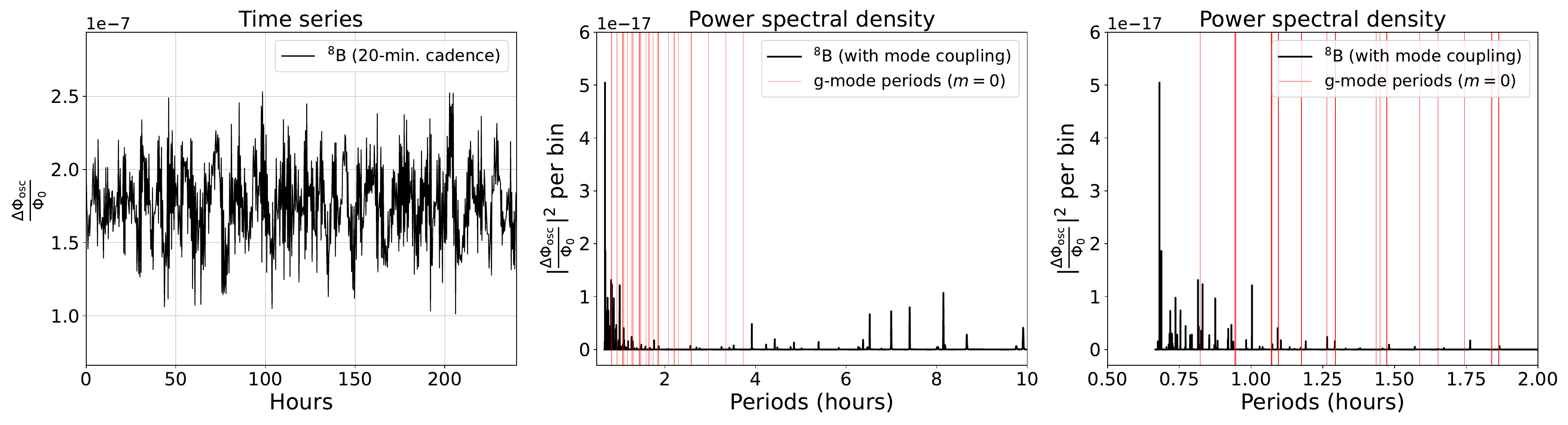}
\caption{\footnotesize Results of numerical evaluation of Equation~(\ref{eq24}) in which about $100$ g-modes~($-8 \le n \le-1$, $\ell=1,\, 2,\, 3$, $-\ell \le m \le +\ell $) are assumed to be contributing to the $^8\mathrm{B}$ neutrino flux fluctuations. 
For the numerical evaluation, the value of $A_{n \ell}$ is assumed to be $10^{-5}$. 
The left panel shows the relative flux fluctuation $\Delta \Phi_{\mathrm{g\text{-}mode}} / \Phi_0$ as a function of time, 
and the middle panel shows the power spectral density of $\Delta \Phi_{\mathrm{g\text{-}mode}} / \Phi_0$ as a function of the period. 
The right panel is a close look into the low-period region~($< 2$ hours) of the middle panel. 
Red lines indicate the periods of $m=0$ g-modes that are used in this analysis. 
It is clearly seen that the peaks in the periodogram (black lines in the middle or right panels) do not necessarily correspond to the individual g-mode periods~(red lines in the middle or right panels), which originate from the mode couplings. 
}

\label{fig:4}
\end{center}
\end{figure}

The leftmost panel of Figure \ref{fig:4} shows an example of the relative fluctuations in the neutrino flux $\Delta \Phi_{\mathrm{g\text{-}mode}} / \Phi_0$ thus evaluated 
in the case of $^{8}\mathrm{B}$ neutrino. 
We see a large number of periodicity in the flux fluctuation, which can be confirmed by the power spectral density of the flux fluctuation 
(the right panels of Figure \ref{fig:4}, where the rightmost panel is a close look into the middle panel). 
However, the fluctuation amplitudes are small ($<  10^{-8}$), and moreover, 
most of the peaks in the periodogram correspond to the periodicity caused by the mode couplings (see the time-varying components in Equation (\ref{eq24})). 
Thus, there are no simple one-to-one relation between the individual g-mode periods (indicated by the red lines in Figure \ref{fig:4}) and the periodicity in the flux fluctuation.

Another point worth mentioning is that 
the temporal average of the relative fluctuation $\Delta \Phi_{\mathrm{g\text{-}mode}} / \Phi_0$ is not zero because of 
the non-time-varying component (see the first term in Equation~(\ref{eq24})). 
The increase in the background neutrino flux of the order of $10^{-7}$ is larger than the amplitudes of the time-varying components~($< 10^{-8}$). 
Importantly, this increase is consistent with the number of the g-modes considered in this subsection~($\sim 100$); when $A_{n \ell}$ is assumed to be $10^{-5}$, \textit{each g-mode contributes to an increase} in the background neutrino flux by $10^{-9}$ (in the case of $^8\mathrm{B}$ neutrinos) as we discussed in Section~\ref{sec:3-3-1}, which amounts to $100 \times 10^{-9} \sim 10^{-7}$ in total. 
This simple proportionality may cause a significant increase in the background neutrino flux as we will discuss in the next subsection.

\subsubsection{Possible long-period neutrino flux variation \label{sec:3-3-3}}

The discussions we had in the previous subsections~(Section~\ref{sec:3-3-1} and Section~\ref{sec:3-3-2}) indicate that we have little chance to detect individual solar g-modes via solar neutrino measurements. 
However, a cumulative second-order effect of a large number of g-modes on the average neutrino flux can be significant, possibly leading to year-long variations in the neutrino flux. 
We will explain this speculative possibility in the following paragraphs. 

We first describe to what extent the second-order neutrino flux fluctuations can affect the average neutrino flux. 
As we discussed in Section~\ref{sec:3-3-2}, the non-time-varying component in the neutrino flux fluctuation ($ \pi \int Q_{(n \ell),(n,\ell)} \mathrm{d}r$) is always positive for the low-degree and low-order modes we analyzed in Section~\ref{sec:3-3-2} (see also Figure \ref{fig:3}), leading to the net increase in the average neutrino flux. 
What if $\pi \int Q_{(n \ell),(n,\ell)} \mathrm{d}r$ is positive for \textit{all} g-modes? 
If this is the case, the net increase in the average neutrino flux might be above the detection limits of future neutrino detectors.
For example, when the number of \textquotedblleft{}involved\textquotedblright{} g-modes is $10^{5}$ and $A_{n \ell}$ is assumed to be constant~(${\sim} 10^{-5}$), the neutrino flux can increase up to $10^{-9} \times 10^{5} \sim 10^{-4}$ in the relative difference (where $10^{-9}$ comes from a contribution from a g-mode in the case $^8\mathrm{B}$ neutrinos). 
An essential point is whether or not a contribution of a g-mode to the non-time-varying component is always positive. 
Actually, this is confirmed to be mostly true based on asymptotic analyses of g-modes~(see Appendix~\ref{App:B}). 

The simple proportionality between the number of g-modes and the net increase in the average neutrino flux leads us to some speculation about 
the possibility that the non-time-varying component can vary with a period as long as the solar cycle ${\sim} 11$~years. 
This may sound weird, as there seems to be no connection between the solar magnetic activity, which mainly is a surface phenomenon, and the solar neutrino, which is coming from the core region. 
However, these two can be interrelated via convection.
The key is the assumption we implicitly adopted that the amplitude parameter $A_{n \ell}$ is constant. 
Since the amplitude parameter is determined by the excitation mechanism of solar g-modes where turbulent convection is thought to play a dominant role~\citep{Gough1985,Kumar+1996,Belkacem+2009,Belkacem+2022}, 
the assumption of the constant $A_{n \ell}$ is valid if the balance between the excitation and damping is frozen. 
Nevertheless, the balance could change in accordance with the solar cycle, resulting in amplitude variations, as has been confirmed in the case of the solar p-modes where the damping rate especially is affected~\citep{Kiefer+Broomhall2021,Howe+2022}. 
This may be especially the case for low-order g-modes; since it is expected that excitation mechanisms of low-order g-modes are similar to those of p-modes~\citep[e.g.,][]{Goldreich+1977,Belkacem+2008,Belkacem+2009,Belkacem+2022}, the amplitudes of low-order g-modes could change periodically with the solar cycle. 
It should also be mentioned that the convective overshooting may have significant roles in g-mode excitation \citep{Pincon2021conf,Pincon+2021}; thus, the solar-cycle variation in physics around the bottom of the convective envelope can affect the amplitude parameter as well.
If such solar-cycle variation in the g-mode excitation is really the case~(although we have to be careful that dominant excitation mechanisms are not completely clear yet even for p-modes as suggested by~\citet{Panetier+2024} and \citet{Panetier+2025}), the non-time-varying component, which is proportional to $A_{n \ell}^2$, also varies with the solar cycle. 
Detections of such a long-period variation in the neutrino flux is relevant, because they might thus be evidence of a bunch of~(low-order) g-modes.

It should finally be noted that we are focusing on the neutrino flux variations caused by solar g-modes in a timescale much shorter than the thermal timescale~($\sim 10^6 -10^7$ years in the case of the Sun). 
In the thermal timescale, the change in the reaction rate~$\varepsilon$ that originates from solar g-mode oscillations can lead to resettlement toward a new equilibrium structure, i.e., the temperature and density profiles would be modified. 
In this respect, our argument could be rephrased as: in timescales much shorter than the thermal timescale, the temporal average of the neutrino flux~(which is fluctuating due to solar g-modes) is always higher than the neutrino flux~($\phi_0$) expected from the temporal averages of the temperature~($T_0$) and density~($\rho_0$), which is mainly due to the strong sensitivity of the reaction rate~$\varepsilon$ especially against the temperature, namely, $\varepsilon \propto T^{\beta}$.

\section{Discussions on the possibility of solar g-mode detection via solar neutrino measurements} \label{sec:4}
In this section, based on the theoretical evaluation in Section~\ref{sec:3}, we discuss the possibility of solar g-mode detection via solar neutrino measurements from observational perspectives. 
We first discuss the current observations in Section~\ref{sec:neutino}, and we then present future prospects in Section~\ref{sec:neutino2}. 

\subsection{Search for periodic signals by analyzing the solar neutrino measurements} \label{sec:neutino}

In this section, we first briefly summarize the experimental searches for the periodic signal by neutrino detectors. As we stated, the $pp$~fusion chain has time scales of order~$10^{4}$~years or more. If the interaction rates at the neutrino detectors vary on any time scale that can be measured, something other than astrophysical phenomena is involved. We focus on relatively short-period variability~($<$ days) in the neutrino flux measurements in Section~\ref{sec:neutrino1}, and we subsequently discuss the longer-period variability~($\sim$ years) in Section~\ref{sec:neutrino2}. Then, in Section~\ref{sec:neutrino3}, 
we discuss the second-order effect to put a constraint on the number of g-mode oscillations inside the Sun by evaluating the amplitude of the solar neutrino flux.

\subsubsection{Experimental search for periodic change with g-mode periodicity} \label{sec:neutrino1}

It is difficult for neutrino detectors to search for high-frequency variations of solar neutrino signals because of the low interaction rate and their background rate. Despite such difficulties, the SNO experiment searched for a periodic change of solar neutrino interaction rate, whose frequency is close to the solar g-mode oscillation, in its solar $\mathrm{^{8}B}$ neutrino sample~\citep{2010ApJ...710..540A}. The collaboration searched for the periodic signal by utilising the Rayleigh power method~\citep{1994MNRAS.268..709B} with the frequencies ranging from $1~\mathrm{day^{-1}}$ to $144~\mathrm{day^{-1}}$. Based on the analysis result, the upper limit on the amplitude of solar $\mathrm{^{8}B}$ neutrinos was set about $11\%$~($90\%$~confidence level~(C.L.)) around the periodicity of solar g-modes. Therefore, no periodic variation with a period of a few hours has so far been established in solar-neutrino data. Borexino also performed detailed temporal analyses of the $\mathrm{^{7}Be}$ neutrino rate using sinusoidal, Lomb--Scargle, and empirical-mode-decomposition methods and found no evidence for periodicities other than the expected annual modulation~\citep{2017APh....92...21A}. To our knowledge, however, apart from the SNO analysis, no published experimental analysis has specifically targeted the hour-scale frequency range relevant to solar g-modes with a dedicated high-frequency search.

\subsubsection{Long-term variation of solar neutrinos} \label{sec:neutrino2}

The amplitude of the annual modulation of solar neutrino fluxes has also been measured by several neutrino detectors. Based on the recent measurements with real-time detection techniques, the annual variation of solar neutrino fluxes is clearly observed, and this amplitude is consistent with the prediction caused by the Kepler orbital eccentricity of the Earth around the Sun~\citep{2023APh...14502778A, 2024PhRvL.132x1803A}. Those measurements demonstrate that neutrino detectors have consistently observed solar neutrino signals under stable operation for several decades. Several experiments, such as Homestake, SAGE, GALLEX/GNO, Super-Kamiokande, and Borexino, have been operated for more than $10$~years and some discussion has occurred about whether these measurement results correlate with the number of sunspots on the Sun\textquoteright{}s surface or not. 

The Homestake experiment reported that the capture rate of solar neutrinos with $\mathrm{^{37}Cl}$ is anti-correlated with the number of sunspots at the surface of the Sun around the solar cycle~21~\citep{Davis:1994jw}. This fact suggested that the neutrino has a finite magnetic moment and the fluctuation of its capture rate is influenced by the time variation of the magnetic field inside the Sun~\citep{1971Ap&SS..10...87C, 1986ZhETF..91..754V}. However, this correlation is not significant at a definitive level~\citep{Bahcall:1987xt} and such scenario is basically excluded because of the small neutrino's magnetic moment~\citep{Beda:2012zz, Borexino:2017fbd, Super-Kamiokande:2020frs, XENON:2022ltv}. A later statistical reanalysis of the complete Homestake data set found only a weak anti-correlation between the $\mathrm{^{37}Cl}$ capture rate and the sunspot number, with a Spearman correlation coefficient of $r_{s}=-0.105$ and a null probability of $14\%$, corresponding to a significance of only approximately $1\sigma$~\citep{2000ApJ...537.1080B}.

The Super-Kamiokande~(SK) data sample contains a lot of interactions and its period ranges from 1996 to 2018. This long-term observation covers nearly two solar activity cycles, i.e., cycles 23 and 24. Based on the measurement, no correlation between the measured yearly solar $\mathrm{^{8}B}$ neutrino flux and the number of sunspots is observed within its uncertainty~\citep{2024PhRvD.109i2001A}. Hence, no periodic signal depending on solar activity is observed. \cite{2024EPJC...84..487P} independently analyzed the SK's solar neutrino data sample with four different statistical methods as well and reached the same conclusion.

\subsubsection{Comparison of the theoretical prediction with the current solar neutrino observations} \label{sec:neutrino3}

In this subsection, we especially focus on year-long variations in the neutrino fluxes, and we put constraints on the number of g-modes by comparing the theoretical evaluation given in Section~\ref{sec:3} with the public data of solar neutrino measurements. In order to estimate the amplitude of neutrino flux with 11-year periodicity, we defined the combination of two sine curves with two different periodicities of 1 year~(annual modulation) and 11 years~(correlated with solar cycle if it exists):

\begin{equation}
    \Phi(t) =  A_{\mathrm{ann}} \sin \left(\omega _{\mathrm{ann}} t +\delta_{\mathrm{ann}} \right)+A_{\mathrm{cycle}} \sin \left(\omega_{\mathrm{cycle}} t +\delta_{\mathrm{cycle}} \right) + C, \label{eq:fit}
\end{equation}

\noindent where $\Phi(t)$ is the solar neutrino flux in neutrino detectors in unit of $\times 10^{6}~\mathrm{cm^{-2} \, s^{-1}}$ for $\mathrm{^{8}B}$ neutrinos, and $\times 10^{9}~\mathrm{cm^{-2} \, s^{-1}}$ for $\mathrm{^{7}Be}$ neutrinos, $A_{\mathrm{ann}}$ is the amplitude of annual modulation caused by the eccentricity of the Earth around the Sun, $A_{\mathrm{cycle}}$ is the amplitude correlated with the solar cycle, $\omega_{\mathrm{ann}}$~($=0.0027~\mathrm{day^{-1}}$) and  $\omega_{\mathrm{cycle}}$~($=2.5\times10^{-4}~\mathrm{day^{-1}}$) are the frequencies of two periodicities, $\delta_{\mathrm{ann}}$ and $\delta_{\mathrm{cycle}}$ are offset times for each periodicity, and $C$ is the constant of the solar neutrino flux, respectively. We should note that their public data contains solar neutrino information with different data formats, such as 5-day flux results by SK, and count rate after subtracting the estimated background by Homestake, SAGE, GALLEX/GNO, and Borexino. According to their data, we calculated the total solar neutrino flux considering the neutrino oscillations to estimate the amplitude of the solar neutrino flux. The details of the calculation for each experiment are described in Appendix~\ref{sec:nu-calc}. Table~\ref{tb:fit} summarizes the resulting parameters from the fitting with Equation~(\ref{eq:fit}) considering the neutrino oscillations. Based on the fitting results, we set upper limits of the parameter~$A_{\mathrm{cycle}}$ as $A_{\mathrm{cycle}}^{\mathrm{Homestake}}< 1.07\times 10^{6}~\mathrm{cm^{-2}} \, \mathrm{s}^{-1}$, and $A_{\mathrm{cycle}}^{\mathrm{SK}}< 0.07\times 10^{6}~\mathrm{cm^{-2}} \, \mathrm{s}^{-1}$ for $\mathrm{^{8}B}$ solar neutrinos, and $A_{\mathrm{cycle}}^{\mathrm{SAGE}}< 0.83\times 10^{9}~\mathrm{cm^{-2}} \, \mathrm{s}^{-1}$, $A_{\mathrm{cycle}}^{\mathrm{GALLEX/GNO}}< 1.11\times 10^{9}~\mathrm{cm^{-2}} \, \mathrm{s}^{-1}$,  and $A_{\mathrm{cycle}}^{\mathrm{Borexino}}< 0.07\times 10^{9}~\mathrm{cm^{-2}} \, \mathrm{s}^{-1}$ for $\mathrm{^{7}Be}$ solar neutrinos with $90\%$~C.L., respectively. 

To evaluate the 11-year periodic change of solar neutrino flux, we also defined its fluctuation as,

\begin{equation}
    F_{11\text{-year}}=A_{\mathrm{cycle}}/C \times100~[\%]. \label{eq:fraction}
\end{equation}

\noindent Our resultant amplitudes correspond to the upper limits of $F_{11\text{-year}}^{\mathrm{Homestake}}<20.9\%$, $F_{11\text{-year}}^{\mathrm{SK}}<1.29\%$, $F_{11\text{-year}}^{\mathrm{SAGE}}<19.7\%$, $F_{11\text{-year}}^{\mathrm{GALLEX/GNO}}<24.7\%$, and $F_{11\text{-year}}^{\mathrm{Borexino}}<1.3\%$ at $90\%$~C.L. with the periodicity of $11$-years. The Super-Kamiokande and Borexino data provide the most stringent constraints on the 11-year modulation, both at approximately the $1.3\%$~level.

\begin{table}[]
    \begin{center}
    \caption{The summary of the parameters fitted with Eq.~(\ref{eq:fit}) and the estimated amplitudes of the total solar $\mathrm{^{8}B}$ and $\mathrm{^{7}Be}$ neutrinos after considering the neutrino oscillation. The parameter $F_{11\text{-year}}$ is defined in Eq.~(\ref{eq:fraction}) as the ratio of amplitude to the constant flux. The unit of $A_{\mathrm{ann}}$, $A_{\mathrm{cycle}}$, and $C$ is $\times 10^{6}~\mathrm{cm^{-2}} \, \mathrm{s}^{-1}$ for $\mathrm{^{8}B}$ neutrinos, and $\times 10^{9}~\mathrm{cm^{-2}} \, \mathrm{s}^{-1}$ for $\mathrm{^{7}Be}$ neutrinos, respectively. The last column is the resulting $\chi^{2}$ with the number of degree of freedom~(n.d.f).}
        \label{tb:fit}
        \begin{tabular}{ccccccc}
        \hline
        Experiment & Solar neutrino & $A_{\mathrm{ann}}$ & $A_{\mathrm{cycle}}$ & $C$ & $F_{11\text{-year}}$ & $\chi^{2}/\mathrm{n.d.f}$ \\
        \hline
        Homestake & $\mathrm{^{8}B}$ & $1.61\pm0.49$ & $<1.07$ & $5.08\pm0.38$ & $<20.9\%$ & $118.85/103=1.15$\\
        Super-Kamiokande & $\mathrm{^{8}B}$ & $0.130\pm0.036$ & $<0.07$ & $5.216\pm0.025$ & $<1.29\%$ & $1591.11/1338=1.19$ \\
        \hline
        SAGE & $\mathrm{^{7}Be}$ & $0.22\pm0.33$ & $<0.83$ & $4.22\pm0.25$ & $<19.7\%$ & $205.82/163=1.26$\\
        GALLEX/GNO & $\mathrm{^{7}Be}$ & $0.52\pm0.50$ & $<1.11$ & $4.62\pm0.35$ & $<24.7\%$ & $147.37/118=1.25$ \\
        Borexino & $\mathrm{^{7}Be}$ & $0.151\pm0.028$ & $<0.07$ & $4.995\pm0.020$ & $<1.3\%$ & $109.25/112=0.98$\\
        \hline
        \end{tabular}
    \end{center}
\end{table}

To test the robustness of the fit against additional periodic components, we repeated the analysis by adding a sinusoidal component with a period of $154$~days, representative of a Rieger-type periodicity, and a component with a period of $2$~years, representative of a quasi-biennial variation. We considered these additional components both separately and simultaneously, with their amplitudes and phases treated as free parameters. Table~\ref{tb:robust} summarizes the resulting $90\%$~C.L. upper limits on
$F_{11\text{-year}}$. The inclusion of these additional periodicities has only a minor
impact on the inferred 11-year modulation. The largest change is found for Homestake, for which the upper limit changes from $20.9\%$ to $22.7\%$ when both additional components are included. For Super-Kamiokande, SAGE, GALLEX/GNO, and Borexino, the changes are less than approximately $2\%$ relative to the nominal results. We therefore conclude that the constraint on the 11-year modulation is robust against the inclusion of these representative additional periodicities.

\begin{table}[]
    \begin{center}
    \caption{Robustness of the $90\%$~C.L. upper limit on the 11-year modulation~($F_{11\text{-year}}$) against additional periodic components.}
        \label{tb:robust}
        \begin{tabular}{ccccc}
        \hline
         Experiment & Nominal & $154$-days & $2$-years & $154$-days + $2$-years \\ 
        \hline
            Homestake   & $<20.9\%$ & $<21.7\%$ & $<21.9\%$ & $<22.7\%$ \\
            Super-Kamiokande & $<1.29\%$ & $<1.30\%$ & $<1.28\%$ & $<1.28\%$ \\
        \hline
            SAGE        & $<19.7\%$ & $<19.5\%$ & $<19.6\%$ & $<19.4\%$ \\
            GALLEX/GNO  & $<24.7\%$ & $<24.7\%$ & $<24.4\%$ & $<24.4\%$ \\
            Borexino    & $<1.36\%$ & $<1.37\%$ & $<1.35\%$ & $<1.35\%$ \\
        \hline
        \end{tabular}
    \end{center}
\end{table}

As explained in Sections~\ref{sec:3-3-2} and~\ref{sec:3-3-3}, each of the contributions from the second-order neutrino fluctuation due to g-modes is positive, and this results in the increase of the solar neutrino flux depending on the number of solar g-modes existing inside the Sun. 
We may approximate a relation between the net increase in the neutrino flux that is caused by a g-mode and $A_{n \ell}$ by taking the arithmetic mean of the net neutrino flux increase caused by each g-mode (see, e.g., Figures~\ref{fig:B3} and \ref{fig:B4}). 
By assuming that the g-mode amplitude~(or $A_{n\ell}$) changes about $10\%$ with the $11$-year solar cycle~\citep[see, e.g.,][in the case of p-modes]{Howe+2022}, we may also derive a relation between the neutrino flux fluctuation associated with the solar cycle and $A_{n \ell}$: $\Delta \Phi_{\mathrm{g\text{-}mode}} / \Phi_0 \sim 0.8 \times (A_{n \ell})^{2}$ for solar $\mathrm{^{8}B}$ neutrinos and $\Delta \Phi_{\mathrm{g\text{-}mode}} / \Phi_0 \sim 0.16 \times (A_{n \ell})^{2}$ for solar $\mathrm{^{7}Be}$ neutrinos. 

\begin{figure}[t]
    \begin{center}
        \includegraphics[width=0.7\linewidth]{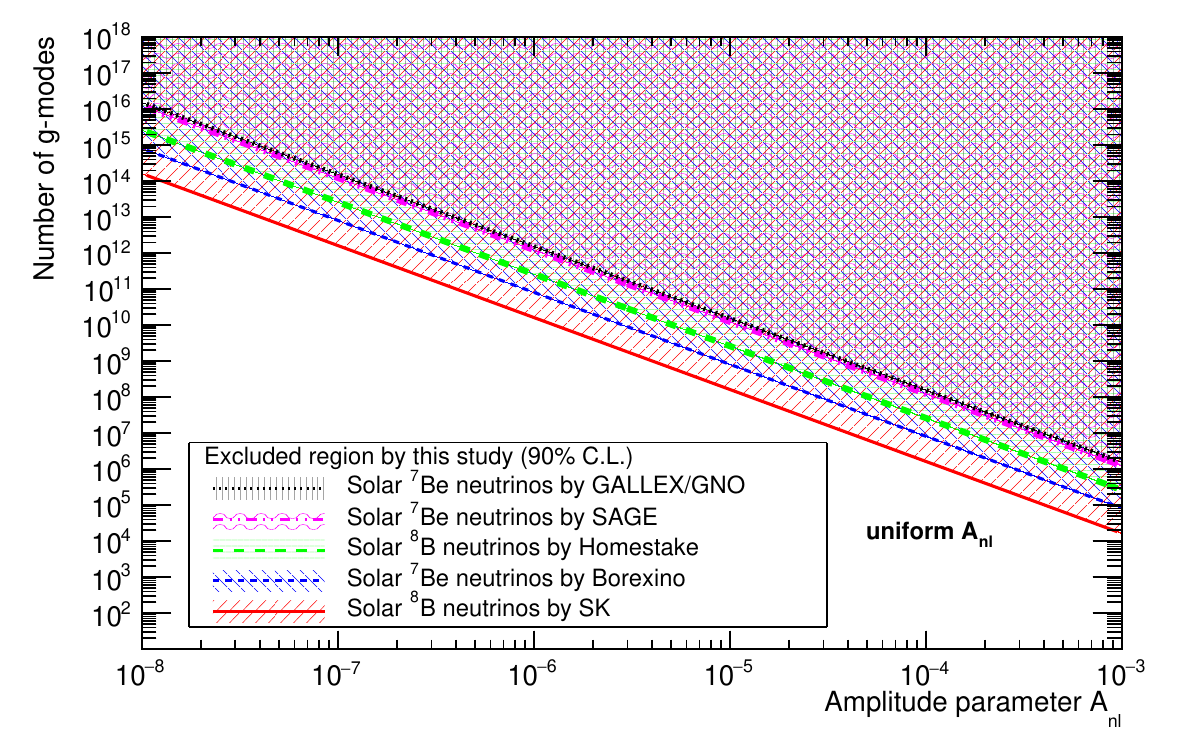}
        \caption{Constraint on the number of g-mode oscillations inside the Sun in the case where we assume $A_{n \ell} = 10^{-5}$ for all g-modes. The thick lines show the upper limit of the number of g-mode oscillations at $90\%$~C.L. by considering the amplitude of the solar $\mathrm{^{8}B}$ neutrinos detected by the Super-Kamiokande detector~(in red), and the Homestake experiment~(in light green), and the solar $\mathrm{^{7}Be}$ neutrinos by the SAGE experiment~(in magenta), and the GALLEX/GNO experiments~(in black), and the Borexino detector~(in blue), respectively. The regions described with diagonal lines show the excluded parameters in this study. The strongest constraint on the number of g-mode oscillations is set by the solar $\mathrm{^{8}B}$ neutrinos measurement by the Super-Kamiokande detector.}
        \label{fig:amplitude}
    \end{center}
\end{figure}

The difference of coefficients basically originates from the difference of the power index~($\eta$) of the reaction rate. Assuming the same amplitude of solar g-mode oscillations $A_{n \ell} \sim 10^{-5}$, the upper limit of the fluctuation with 11~years sets the upper limits~($90\%$~C.L.) of the number of solar g-modes to be $N_{\mathrm{g\text{-}mode}}^{\mathrm{{^{8}B}\,(SK)}}<1.6 \times 10^{8}$ by $\mathrm{^{8}B}$ neutrinos by the Super-Kamiokande detector and $N_{\mathrm{g\text{-}mode}}^{\mathrm{{^{7}Be}\,(Borexino)}}<8.1 \times 10^{8}$ by $\mathrm{^{7}Be}$ neutrinos by the Borexino detector, respectively. Although the Borexino's data~($\mathrm{^{7}Be}$ neutrinos) provides the highest solar neutrino flux measurement accuracy, the Super-Kamiokande's data~($\mathrm{^{8}B}$~neutrinos) sets the strongest constraint on the number of g-mode oscillations inside the Sun because of the difference in the coefficient arising from different temperature dependencies, i.e. the power index of the temperature~$T^{\eta}$. Figure~\ref{fig:amplitude} shows the constraint on the number of g-mode oscillations for the cases with different values of $A_{n \ell}$. Although we estimated the possible number of solar g-mode oscillations in the Sun under several assumptions, we emphasize that this analysis is the first attempt to constrain the number of g-mode oscillations existing in the Sun. 

For comparison, we also would like to present similar analyses where we lift the assumption that all the g-modes have the same amplitude $A_{n \ell}$. 
To this end, we have followed \citet{Belkacem+2009}, and have expressed in an ad hoc manner the dependence of $A_{n \ell}$ on the g-mode frequency $\nu_g$ and the spherical degree $\ell$ as follows: $A_{n \ell} = A_{0} \times (\nu_g / \nu_0)^{2.5} \times ( \ell (\ell + 1)/2 )^{-1}$, where $\nu_0$ indicates a certain g-mode frequency (here set to be $100 \, \mu \mathrm{Hz}$), and $A_{0}$ represents the relative amplitude (basically treated as a free parameter) for the corresponding g-mode. 
In the same manner as we described in the case of uniform $A_{n \ell}$ for all the g-modes, we may derive relations between the solar-cycle variation in the neutrino flux and the amplitude parameter $A_{0}$: $\Delta \Phi_{\mathrm{g\text{-}mode}} / \Phi_0 \sim 0.3 \times (A_{0})^{2}$ for solar $\mathrm{^{8}B}$ neutrinos and $\Delta \Phi_{\mathrm{g\text{-}mode}} / \Phi_0 \sim 0.06 \times (A_{0})^{2}$ for solar $\mathrm{^{7}Be}$ neutrinos (see also Appendix~\ref{App:B}); roughly speaking, there are slight differences between the two kinds of relations that are derived based on the different distributions of $A_{n \ell}$. 
When we assume $A_0=10^{-5}$, the illustrative non-uniform amplitude distribution gives $N_{\rm g\text{-}mode}^{^{8}{\rm B}({\rm SK})}
\lesssim 4.3\times10^{8}$ and $N_{\rm g\text{-}mode}^{^{7}{\rm Be}({\rm Borexino})}
\lesssim 2.2\times10^{9}$, which are approximately a factor of $2.7$~weaker than the
corresponding limits obtained under the uniform-amplitude assumption.

\subsection{Prospects for the g-mode oscillation search with neutrinos} \label{sec:neutino2}
We here discuss the possibility of the solar g-mode detection via solar neutrino measurements by future neutrino detectors. If large-sized detectors, such as Hyper-Kamiokande, DUNE, SNO${+}$ and JUNO, are continuously operated, the precision of solar neutrino flux measurements is improved. Such measurements have a chance to measure small amplitude of fluctuations with long periodicities. We may further set the constraint on the number of g-mode oscillations by the same analysis method once their results are released. 

\subsubsection{Prospect with water Cherenkov detectors}

As we stated, searches for high-frequency variation by neutrino detectors are challenging due to the low interaction rate and background control. However, it is still important to search for relatively short periodic change, corresponding to the g-mode periodicity ($\sim$ a few hours), by neutrino detectors and to verify our theoretical argument that no fluctuation is expected to first-order, as described in Section~\ref{sec:2-2}. Once no fluctuation is confirmed, 
our theoretical formulation is validated. On the other hand, the existence of periodic fluctuation may require further considerations. For example, we in this study assume that the background state is spherically symmetric. 
But the Sun is characterized by the measurable oblateness \citep[$\Delta_\odot \sim 10^{-5}$, where $\Delta_\odot$ represents the relative difference between the equatorial and polar radii;][]{Kuhn+2012,Meftah+2015}, and the background state can be latitudinally dependent. Therefore, the latitudinal integration would not be zero as we discussed in Section \ref{sec:2-2}, and there would be finite first-order neutrino flux fluctuations. 
In addition, nonadiabatic oscillation and stochastic excitation were neglected in this study. 
Although these neglected effects are estimated to be small (Appendix~\ref{sec:4-1-1}), the estimates are based on a typical picture of the solar interior; thus, some loopholes may exist. 
It should also be instructive to mention that a recent numerical study predicts previously unnoticed global modes of the Sun that oscillate with a period of about $1$ hour, and have sensitivity to the solar core \citep{LeSaux+2025}.
 
Hence, the search for relatively short periodic change in solar neutrino flux by neutrino detectors is very helpful for further understanding of the inner structure of the Sun.

The Hyper-Kamiokande experiment is one of the most promising detectors to search for high-frequency variation in solar $\mathrm{^{8}B}$ neutrinos. Once the Hyper-Kamiokande starts its operation in the middle of 2028, more solar $\mathrm{^{8}B}$ neutrino events will be observed because its fiducial volume for the physics analysis is eight times larger than that of the Super-Kamiokande. This larger volume allows detecting about five to ten solar $\mathrm{^{8}B}$ neutrino events per hour, depending on its data acquisition~(energy) threshold and background control. If such measurement is achieved, the HK detector can determine the fluctuation of solar $\mathrm{^{8}B}$ neutrinos with the periodicities of the solar g-mode oscillations, where LT14 predicts. For this reason, we request the Super-Kamiokande and Hyper-Kamiokande collaborations to analyse the observed data according to the analysis method developed by the SNO collaboration~\citep{2010ApJ...710..540A} in order to evaluate the fluctuation of solar $\mathrm{^{8}B}$ neutrinos due to solar g-mode oscillations.

For the search for long periodic change, we also request the Super-Kamiokande collaboration to update their solar neutrino analysis using the data set after 2018, because such data covers the period of the solar cycle~$25$. If the Super-Kamiokande continues to operate until 2030, the SK data will cover three solar cycles, 23, 24, and 25. Such long measurements can set a stronger constraint on the number of solar g-mode oscillations, which is demonstrated in Figure~\ref{fig:amplitude}.

\subsubsection{Prospect with liquid scintillator detectors}

As of 2026, two liquid scintillator detectors, e.g. SNO${+}$~\citep{SNO:2021xpa} and JUNO~\citep{JUNO:2025fpc}, have been operated to observe solar neutrinos in the energy range of keV. The SNO${+}$ detector is located in Canada, with a target mass of $0.78$~ktons of liquid scintillator, and the JUNO detector is located in China,  with a target mass of $20.0$~ktons of liquid scintillator. Both two detectors can detect $pp$, $\mathrm{^{7}B}$, $pep$, $\mathrm{^{8}B}$, and CNO solar neutrinos with their low background rate and high energy resolution~\citep{SNO:2025qqy, JUNO:2025fpc}.

For JUNO, the sensitivity to a periodic variation of the solar $\mathrm{^{7}Be}$ neutrino interaction rate has been evaluated under four background assumptions~\citep{JUNO:2023zty}. After 10 years of exposure, the projected $3\sigma$ sensitivity to the modulation amplitude is about $0.3\%$ in the Very Low background scenario, $0.4\%$ in the Low scenario, $0.8\%$ in the Medium scenario, and $2.5\%$ in the High scenario. Thus, the achievable sensitivity spans approximately $0.3$--$2.5\%$, depending on the residual background level. Assuming that a comparable precision can be achieved for the long-term flux variation considered here, the corresponding constraint on the number of g-mode oscillations is estimated to be

\begin{equation}
    N_{\mathrm{g\text{-}mode}}^{\mathrm{{^{7}Be},(JUNO)}} \sim (2\text{--}16)\times10^{8},
\end{equation}

\noindent where $A_{n\ell}=10^{-5}$ is assumed.

\section{Conclusion} \label{sec:5}

The scientific return from a definitive detection of solar g-modes would be substantial, providing unprecedented constraints on the rotation, composition, and dynamics of the solar core.
The tiny g-mode amplitudes at the solar surface, however, have prevented us from the firm detection of the solar g-modes via optical measurement. 
Based on the theory of linear adiabatic oscillation of a spherically symmetric star, we have assessed the possibility of solar g-mode detection via solar neutrino measurements.  
Our theoretical assessment is mostly negative. 
Firstly, the first-order fluctuation caused by g-modes should be zero due to the geometrical cancellation; 
thus, the first-order formulation, whilst remaining physically valid as a local sensitivity measure, requires extension to second order to predict the total integrated flux variation, as we have shown.

The first-order fluctuation can be non-zero if we take into account the effects of the time-of-flight of the neutrinos from the solar interior to an observational point (time-delay effect). 
Still, even when we consider the time-delay effect, the first-order relative fluctuations in the neutrino flux caused by a g-mode (with the amplitude parameter $A_{n \ell} = 10^{-5}$) is $< 2 \times 10^{-9}$, that is by far smaller than the detection limit of any existing neutrino detectors. 
Secondly, we evaluated the second-order neutrino flux fluctuations caused by a g-mode (with $A_{n \ell} = 10^{-5}$) that is again too small to detect ($\sim 10^{-9}$ and $\sim 10^{-10}$ in relative difference, in the case of $^8\mathrm{B}$ and $^7\mathrm{Be}$ neutrinos, respectively). 
The predicted signal ($\sim 10^{-10}-10^{-9}$) lies approximately seven orders of magnitude below current experimental sensitivity~(SNO upper limit $\sim 11\%$ for periodic variation).

However, our theoretical assessment is not totally negative. 
The second-order analysis also enabled us to find a hint that the non-time-varying component in the neutrino flux fluctuation can lead to a significant increase in the average neutrino flux ($\sim 10^{-5}$ in relative difference, when $A_{n \ell}$ and $N_{\mathrm{g\text{-}mode}}$ are assumed to be $10^{-5}$ and $10^5$, respectively, where $N_{\mathrm{g\text{-}mode}}$ represents the number of g-modes involved). 
Since this increase in the average neutrino flux is closely related to the amplitude parameter $A_{n \ell}$, which is determined by excitation/damping by the turbulent convection and thus could change in accordance with the solar cycle, the average neutrino flux can also change with the solar cycle. 
Therefore, the second-order analysis we conducted in this study suggests that monitoring the solar neutrino flux variations for a long time ($\sim$ a few decades) could be of great help to understand the excitation mechanisms of the solar g-modes. 
As a first step toward such an attempt, we analyzed the public data from the neutrino experiments, gave an upper limit on the amplitude of the solar neutrino variation that has a period of ${\sim} 11$~years, and compared the upper limit with the theoretical assessment conducted in this study, ending up with the upper limit on the number of g-modes inside the Sun $N_{\mathrm{g\text{-}mode}}^{\mathrm{^{8}B\,(SK)}}< 1.6 \times 10^{8}$ at $90\%$~C.L. where $A_{n\ell}$ is assumed to be $10^{-5}$. 
Although our analyses may be preliminary, the attempts are important in the context that we are awaiting future neutrino detectors, which will significantly enrich the solar neutrino observations.

Finally, we note that neutrino energy spectrum measurements can also constrain the amplitude of g-mode oscillations because perturbations of the electron-density profile modify the solar electron-neutrino survival probability through the MSW effect~\citep{1996PhRvD..54.3941B, 1998NuPhB.513..319B, 2003ApJ...588L..65B}. For the fiducial amplitude $A_{n\ell}=10^{-5}$ adopted in this study, which corresponds to a maximum relative temperature perturbation of order $|\delta T/T|\sim10^{-5}$, the adiabatic relation implies a fractional density perturbation, and hence an electron-density perturbation, of order a few $\times10^{-5}$ in the solar core. Linearizing the standard adiabatic MSW survival probability with respect to the electron density then gives an absolute change of $|\delta P_{ee}|\sim10^{-6}$ for solar $^{8}$B neutrinos in the several-MeV energy region. This change is far below the present experimental precision on the solar-neutrino survival probability. A detailed calculation including the spatial structure of the g-modes, the neutrino production profile, and the propagation geometry will be presented in a subsequent study.

\begin{acknowledgments}

We thank K.~Belkacem, G.~Buldgen, and J.~Schou for the insightful discussions. We also thank M.~Nakahata for telling us about the solar neutrino measurements conducted last four decades. We would also like to thank the anonymous referee for the insightful comments. This work is supported by MEXT KAKENHI Grant Numbers: 23KJ0300~(YH), 24K00654~(YN), 24K07099~(MK), and 24K17087~(YH), and JST SPRING Japan Grant Number JPMJSP2178~(SS). Y.H. acknowledges Wilhelm und Else Heraeus-Foundation for the financial support to participate in ``Interdisciplinary Physics of the Sun''.  This work was carried out by the joint research program of the Institute for Space-Earth Environmental Research~(ISEE), Nagoya University. In particular, we thank H.~Hotta of ISEE for the program application and his hospitality during our visit.

\end{acknowledgments}

\bibliography{main}{}
\bibliographystyle{aasjournal}



\appendix
\section{Neglected effects in the derivation of Equation (6)} \label{sec:4-1-1} 
We here discuss some effects that are neglected in deriving Equation (\ref{eq6}). 
After we evaluate the time-delay effect, which actually is found to be dominant among other neglected effects (Appendix \ref{sec:a1}), we give qualitative discussions on the other effects, such as asphericity and nonadiabaticity (Appendix \ref{sec:a2}). 

\subsection{Time-delay effect} \label{sec:a1} 
As we see in Section~\ref{sec:2}, when we neglect the time-delay effect and assume that all neutrinos produced inside the Sun arrive at a certain observational point at the same time, the first-order fluctuation is proportional to the horizontal integration of the spherical harmonics, which results in zero (see Equation~(\ref{eq11})). 
We here would like to demonstrate that there would be non-zero first-order fluctuations if we take into account the time-delay effects, though the effect is smaller than the second-order fluctuations numerically evaluated in Section~\ref{sec:3-3}. 

We first show the first-order fluctuation in the neutrino flux that is caused by g-modes (see also Equation~(\ref{eq6})): 
\begin{eqnarray}
\Delta \Phi_{\mathrm{g\text{-}mode}} = \int [\phi_0 \rho' + \rho_0 \phi'] \mathrm{d} V. \label{eq6_2}
\end{eqnarray}

\noindent If we consider the time-delay effects, the equation above can be rewritten in the following form:

\begin{eqnarray}
\Delta \Phi_{\mathrm{g\text{-}mode}}(t) 
= 
\int [\phi_0(r) \rho'(r,\theta,\psi,t - d/c) + \rho_0(r) \phi'(r, \theta, \psi, t - d/c)] \mathrm{d} V, \label{eq27}
\end{eqnarray}

\noindent where $t$ is the time at the observational point, and $d$ is the distance between the observational point and a certain point inside the Sun designated by $(r,\theta, \psi)$. 
It is assumed that neutrinos propagate with the light speed~$c$. 

In almost the same manner as described in Section~\ref{sec:2-2}, we can analyze Equation~(\ref{eq27}). 
One exception is in the functional forms of the eigenfunctions~(Equations~(\ref{eq9}) and (\ref{eq10})) where $\mathrm{cos}(m\psi - \omega_{n \ell m}t)$ is replaced by $\mathrm{cos}(m\psi - \omega_{n \ell m}(t - d / c))$. 
By introducing a constant $d_0$, that is the distance between the observational point and the solar center, $d$ can be expressed as $d(r,\theta, \psi) = d_{0} + \delta d(r, \theta, \psi)$, with which we have: 

\begin{eqnarray}
\mathrm{cos}(m\psi - \omega_{n \ell m}(t - d / c)) 
& = & 
\mathrm{cos}(m\psi - \omega_{n \ell m} t - \delta'_{n \ell m}) 
\mathrm{cos}(\omega_{n \ell m} \delta d / c) 
- 
\mathrm{sin}(m\psi - \omega_{n \ell m} t - \delta'_{n \ell m}) 
\mathrm{sin}(\omega_{n \ell m} \delta d / c) \nonumber \\ 
& \sim & 
\mathrm{cos}(m\psi - \omega_{n \ell m} t - \delta'_{n \ell m}) 
- 
\mathrm{sin}(m\psi - \omega_{n \ell m} t - \delta'_{n \ell m}) 
(\omega_{n \ell m} \delta d / c). \label{eq28}
\end{eqnarray}

\noindent The extra phase $\delta_{n \ell m}'$ is defined as $\delta_{n \ell m}' = - \omega_{n \ell m} (d_0 / c)$. 
The last line is obtained by the first-order Taylor expansion for the trigonometric function, which is reasonable because the time-difference ($\delta d / c \sim 0.1 \, R_{\odot} / c$) is of the order of $10^{-1}$ seconds while a possible shortest g-mode period ($\propto \omega_{n \ell m}^{-1}$) is $\sim 2 \times 10^{3}$ seconds (see also Figure \ref{fig:N2} for the highest value of the BV frequency). 

As we discussed in Section~\ref{sec:2-2}, the terms that include the first term of the last line in Equation~(\ref{eq28}) are zero due to the geometrical cancellation. 
Therefore, what are left in the first-order neutrino flux fluctuation are those including the second term of the last line in Equation~(\ref{eq28}). 
Since $d \gg r$ and $d \gg \delta d$, we can approximate $\delta d \sim - r \mathrm{cos} \alpha$, in which $\alpha$ is the angle between a vector pointing toward the observational point from the solar center and a position vector $r \boldsymbol{e}_{r}$. 
If we assume that the equatorial plane in the heliographic coordinate is parallel to the line-of-sight vector, $\mathrm{cos} \alpha = \mathrm{sin} \theta \mathrm{cos} \psi$, allowing us to analytically carry out the integration over the latitude and the azimuth. 
We thus end up with the following expression for the first-order fluctuation in the neutrino flux where the time-delay effect is taken into account: 

\begin{eqnarray}
\Delta \Phi_{\mathrm{g\text{-}mode}}(t) 
= 
\frac{2 \pi }{\sqrt{3}} 
\sum_{n} 
\frac{\omega_{n 1 1}}{c} 
\int 
G_{n1}(r) r 
\mathrm{d}r 
\times 
\mathrm{sin}(\omega_{n 1 1} t + \delta'_{n 1 1}) 
- 
\frac{2 \pi }{\sqrt{3}} 
\sum_{n} 
\frac{\omega_{n 1 -1}}{c} 
\int 
G_{n1}(r) r 
\mathrm{d}r 
\times 
\mathrm{sin}(\omega_{n 1,-1} t + \delta'_{n 1,-1}). \nonumber \\ 
\label{eq29}
\end{eqnarray}

\noindent See Equation~(\ref{eq12}) for the definition of the function ~$G_{n \ell}$. Due to the angular dependence of $\mathrm{sin} \theta \mathrm{cos} \psi$, which is newly introduced via the approximation of $\delta d$, only dipole modes with $m = \pm 1$ remain in the first-order fluctuation above. 

Using the same models and eigenfunctions described in Sections~\ref{sec:3-1} and~\ref{sec:3-2}, we numerically evaluated the amplitudes of the first-order time-varying component in Equation (\ref{eq29}). 
When the amplitude parameter~$A_{n \ell}$ is assumed to be $10^{-5}$, the maximum of the relative difference $\Delta \Phi_{\mathrm{g\text{-}mode}} / \Phi_0$ thus evaluated are $< 2 \times 10^{-9}$. 
The first-order fluctuation where the time-delay effect is taken into account is therefore comparable to or smaller than the second-order fluctuation~($\sim 10^{-9}$), rendering the time-delay effects not to be significant in our analyses shown in Section~\ref{sec:3-3}. 

Despite the small contribution to the neutrino flux fluctuations, it might be instructive and interesting to mention that the time-delay effects have an angular dependence on the solar inclination angle~\citep[e.g.][]{Gizon+1998,Gizon+Solanki2003}; that is, if the pulsation axis of the Sun is not perpendicular to the line-of-sight vector as was assumed in the discussions above, integration over horizontal direction leads to results different from Equation~(\ref{eq29}). 
Therefore, we may see first-order fluctuations caused by g-modes other than the dipole modes with $m = \pm 1$. 
Note again that the chance to observe the signature is quite small due to the fairly small time-delay effects. 

\subsection{Other possibilities} \label{sec:a2}
Other than the time-delay effect, any deviation from our assumption, such as the aspherical structure and nonadiabatic oscillation, can lead to ``residuals'' in the first-order neutrino flux fluctuation. 
We here aim at giving some qualitative assessments on these effects neglected in derivation of Equation (\ref{eq6}). 

We start with the asphericity. 
From the observational perspective, it is well known that the Sun is oblate, and the relative difference between the equatorial and polar radii has been estimated of the order of $10^{-5}$~\citep[e.g.,][]{Kuhn+2012,Meftah+2015}. 
Such asphericity has been believed to persist throughout the solar interior, which is quite natural as is evidenced by the existence of differential rotation in the convective envelope, though it has been predicted to be fairly small; for example, direct inferences on the aspherical structures (e.g., the sound speed profile) have not been converged yet at the moment~\citep[e.g.,][]{Antia+2001,Brun+2010}. 
Numerical studies of the solar inertial modes indicate that the temperature relative difference between the equator and the pole can be $7 \, \mathrm{K}$ in the middle of the convection zone~\citep{Bekki+2024}. 

Importantly, the aspherical properties can render the horizontal integration in, e.g., Equation (\ref{eq8}), to be nonzero, leading to residuals of the first-order neutrino flux fluctuation. 
In particular, remembering that the first-order neutrino flux fluctuation is determined by the product between the temperature perturbation $\delta T_{n \ell}$ and the equilibrium neutrino flux $\phi$ (see the bottom row of Figure \ref{fig:2}), the asphericity around the solar core region matters. 
When we assume that the degree of the local asphericity is given by the ratio between the centrifugal force and the local gravity, the ratio is proportional to $r^{3}$; thus, around the core region ($r/R_{\odot} \sim 0.1$), the asphericity can be estimated to be of the order of $10^{-8}$. 
The radial integration of $G_{n \ell}(r)$ (Equation (\ref{eq12})) is of the order of $1$. 
Consequently, the residual first-order neutrino flux fluctuation is around $10^{-13}$ (we assume $A_{n \ell} = 10^{-5}$), which is much smaller than the time-delay effect and the second-order fluctuations in the neutrino flux.

We then move on to discussing nonadibatic oscillation, which can be relevant especially for g-modes whose periods can be comparable to the local thermal timescale somewhere inside the Sun (mainly very surface regions). 
The nonadiabaticity can shift the phase of the g-mode radial eigenfunctions. 
Still, it should be mentioned that, if the equilibrium structure is spherically symmetric, the horizontal integration is zero even in the case of nonadiabatic oscillation. 
In other words, we need both asphericity and nonadiabatic oscillation to produce the residual first-order neutrino flux fluctuation. 
As estimated in the last paragraph, the asphericity can cause just $10^{-13}$ residuals, and thus, the combined effects of asphericity and nonadiabaticity should be much smaller. 
We also would like to mention that the oscillation is almost adiabatic around the solar core region (see discussions in Section \ref{sec:2-1}).

We finally would like to remark that in our derivation of Equation (\ref{eq6}) we assume that the g-mode lifetimes are infinite, which is actually not the case because the g-modes are considered to be excited by turbulent convection \citep{Gough1985,Belkacem+2009}. 
Stochastic excitation can cause larger root-mean-square velocities, but again, the horizontal integration is zero if the equilibrium structure is spherically symmetric. 

Overall, the most dominant contribution to the residual first-order neutrino flux fluctuation comes from the time-delay effect, which is smaller than or comparable to the second-order fluctuations.

\section{Notes on how to derive Equation (20)} \label{App:A}
We here present details on derivation of Equation (\ref{eq24}), where the function $Q_{(n\ell),(n' \ell')}$ is defined as a sum of functions as below: 
\begin{eqnarray}
Q_{(n\ell),(n' \ell')} = \sum_{i=1}^{4} Q^i_{(n \ell),(n' \ell')}, \label{eq25}
\end{eqnarray}
for which $Q^i_{(n\ell),(n' \ell')}$ is defined as: 
\begin{eqnarray}
Q^1_{(n\ell),(n' \ell')} 
&=& [c_1 (\Gamma_{3,0} - 1)^{-1} + c_2 ] 
\biggl ( \frac{\delta T_{n\ell}}{T_0} \biggr ) 
\biggl ( \frac{\delta T_{n' \ell'}}{T_0} \biggr ) 
\phi_0 \rho_0 r^2, \label{eq25_1} \\
Q^2_{(n\ell),(n' \ell')} 
&=& 
-(\Gamma_{3,0} - 1)^{-1} 
\biggl ( \frac{\delta T_{n \ell}}{T_0} \biggr )
\xi_{r,n' \ell'} 
\biggl ( 
\frac{\partial \, \mathrm{ln} \, \phi_{0}}{\partial r} 
\biggr )
\phi_0 \rho_0 r^2, \label{eq25_2} \\
Q^3_{(n\ell),(n' \ell')} 
&=& 
-c_1
\biggl ( \frac{\delta T_{n \ell}}{T_0} \biggr )
\xi_{r,n' \ell'} 
\biggl ( 
\frac{\partial \, \mathrm{ln} \, \rho_{0}}{\partial r} 
\biggr ) 
\phi_0 \rho_0 r^2, \label{eq25_3} \\
Q^4_{(n\ell),(n' \ell')} 
&=& 
\xi_{r,n \ell} \xi_{r,n' \ell'} 
\biggl ( 
\frac{\partial \, \mathrm{ln} \, \phi_{0}}{\partial r} 
\biggr ) 
\biggr ( 
\frac{\partial \, \mathrm{ln} \, \rho_{0}}{\partial r} 
\biggr ) 
\phi_0 \rho_0 r^2. \label{eq25_4} 
\end{eqnarray} 
Physically speaking, $Q^{1}$ represents the direct temperature-temperature coupling, $Q^2$ and $Q^3$ the cross-terms between temperature perturbation and radial displacement, and $Q^4$ the displacement-displacement coupling.
 
We start with the computation of the first term in the right-hand side of Equation (\ref{eq19}) 
for which Equations (\ref{eq16}) and (\ref{eq20}) are substituted, leading to the following expression: 
\begin{eqnarray}
(\mathrm{1st}) & = & 
\sum_{n \ell m} 
\sum_{n' \ell'm'} 
\int_{0}^{R_{\star}} 
[c_1 (\Gamma_3 - 1)^{-1} + c_2]
\biggl ( \frac{\delta T_{n \ell}}{T_0} \biggr )
\biggl ( \frac{\delta T_{n' \ell'}}{T_0} \biggr )
\phi_0 \rho_0 r^2 
\mathrm{d}r
\times \int_{-1}^{1} P^{m}_{\ell}(\mu) P^{m'}_{\ell'}(\mu) 
\mathrm{d} \mu \nonumber \\ 
& \times & 
\biggl \{ 
\int_{0}^{2 \pi} 
\mathrm{cos} \, (m\psi) 
\mathrm{cos} \, (m'\psi) 
\mathrm{d} \psi \times 
\mathrm{cos} \, (o(t)) 
\mathrm{cos} \, (o'(t)) 
+ 
\int_{0}^{2 \pi} 
\mathrm{sin} \, (m\psi) 
\mathrm{sin} \, (m'\psi) 
\mathrm{d} \psi \times 
\mathrm{sin} \, (o(t)) 
\mathrm{sin} \, (o'(t)) 
\biggr \},  
\label{eq_A1}
\end{eqnarray}
where $\mu = \mathrm{cos} \, \theta$. 
The definitions of the other variables can be found in the main text. 

Due to the orthogonality of the trigonometric functions, 
the azimuthal integration is zero unless $m' = \pm m$, resulting in: 
\begin{eqnarray}
(\mathrm{1st}) & = & 
\sum_{(n,\ell,m \neq0)} 
\sum_{(n',\ell',m'=m)} 
\int_{0}^{R_{\star}} 
Q^{1}_{(n \ell),(n' \ell')} 
\mathrm{d}r
\times 
\int_{-1}^{1} P^{m}_{\ell}(\mu) P^{m}_{\ell'}(\mu) \mathrm{d} \mu 
\times 
\biggl \{ 
I_{\mathrm{c},m} 
\times 
\mathrm{cos} \, (o) 
\mathrm{cos} \, (o') 
+ 
I_{\mathrm{s},m} 
\times 
\mathrm{sin} \, (o) 
\mathrm{sin} \, (o') 
\biggr \} \nonumber \\ 
& + & 
\sum_{(n,\ell,m\neq0)} 
\sum_{(n',\ell',m'=-m)} 
\int_{0}^{R_{\star}} 
Q^{1}_{(n\ell),(n'\ell')} 
\mathrm{d}r
\times \int_{-1}^{1} P^{m}_{\ell}(\mu) P^{-m}_{\ell'}(\mu) \mathrm{d} \mu 
\times 
\biggl \{ 
I_{\mathrm{c},m} 
\times 
\mathrm{cos} \, (o) 
\mathrm{cos} \, (o') 
- 
I_{\mathrm{s},m} 
\times 
\mathrm{sin} \, (o) 
\mathrm{sin} \, (o') 
\biggr \} \nonumber \\
& + & 
\sum_{(n,\ell,m=0)} 
\sum_{(n',\ell',m'=0)} 
\int_{0}^{R_{\star}} 
Q^{1}_{(n \ell),(n' \ell')} 
\mathrm{d}r
\times \int_{-1}^{1} P^{0}_{\ell}(\mu) P^{0}_{\ell'}(\mu) \mathrm{d} \mu 
\times 
\biggl \{ 
I_{\mathrm{c},0} 
\times 
\mathrm{cos} \, (o) 
\mathrm{cos} \, (o') 
+ 
I_{\mathrm{s},0} 
\times 
\mathrm{sin} \, (o) 
\mathrm{sin} \, (o') 
\biggr \}, \nonumber \\  
\label{eq_A2}
\end{eqnarray}

\noindent 
where $Q_{(n \ell),(n' \ell')}^{1}$ is defined in Equation (\ref{eq25_1}). 
The azimuthal integrations, namely, 
$\int_{0}^{2 \pi} \mathrm{cos}^2(m\psi) \mathrm{d}\psi$ and 
$\int_{0}^{2 \pi} \mathrm{sin}^2(m\psi) \mathrm{d}\psi$, 
are denoted by $I_{\mathrm{c},m}$ and $I_{\mathrm{s},m}$, respectively. 

We then carry out the latitudinal integration in Equation (\ref{eq_A2}). 
We hereafter consider that the associated Legendre polynomials are normalized so that their square integrals are unity. 
We thus have a relation that 
$P_{\ell}^{-m} = (-1)^m P_{\ell}^{m}$. 
In addition, by focusing on the orthogonality of $P_{\ell}^{m}$ in terms of a fixed $m$, 
the latitudinal integration is zero unless $\ell' = \ell$, and therefore, we have: 
\begin{eqnarray}
(\mathrm{1st}) & = & 
\sum_{(n,\ell, m\neq0)} 
\sum_{(n',\ell'=\ell,m'=m)} 
\pi 
\int_{0}^{R_{\star}} 
Q^{1}_{(n \ell),(n' \ell')} 
\mathrm{d}r 
\times 
\mathrm{cos} \, (o - o') \nonumber \\ 
& + & 
\sum_{(n,\ell,m\neq0)} 
\sum_{(n',\ell'=\ell,m'=-m)} 
\pi (-1)^m 
\int_{0}^{R_{\star}} 
Q^{1}_{(n \ell),(n' \ell')} 
\mathrm{d}r
\times 
\mathrm{cos} \, (o + o')  \nonumber \\
& + & 
\sum_{(n,\ell,m=0)} 
\sum_{(n',\ell'=\ell,m'=0)} 
2\pi 
\int_{0}^{R_{\star}} 
Q^{1}_{(n \ell),(n'\ell')} 
\mathrm{d}r 
\times 
\mathrm{cos} \, (o) 
\mathrm{cos} \, (o'). \nonumber \\  
\label{eq_A3}
\end{eqnarray}
Note that $I_{\mathrm{c},m} = I_{\mathrm{s},m} = \pi$ for non-zero $m$, 
and that $I_{\mathrm{c},0} = 2 \pi$ and $I_{\mathrm{s},0} = 0$ for $m=0$. 
Finally, to explicitly see the non-time-varying component, 
we extract terms with $n = n'$, with which $\mathrm{cos}\, (o - o')$ becomes unity, from the first term of the right-hand side in Equation~(\ref{eq_A3}), enabling us to rewrite the equation as below: 
\begin{eqnarray}
(\mathrm{1st}) & = & 
\sum_{(n,\ell, m\neq0)} 
\sum_{(n'=n,\ell'=\ell,m'=m)} 
\pi 
\int_{0}^{R_{\star}} 
Q^{1}_{(n \ell),(n' \ell')} 
\mathrm{d}r \nonumber \\ 
& + &
\sum_{(n,\ell, m\neq0)} 
\sum_{(n'\neq n,\ell'=\ell,m'=m)} 
\pi 
\int_{0}^{R_{\star}} 
Q^{1}_{(n \ell),(n' \ell')} 
\mathrm{d}r 
\times 
\mathrm{cos} \, (o - o') \nonumber \\ 
& + & 
\sum_{(n,\ell,m\neq0)} 
\sum_{(n',\ell'=\ell,m'=-m)} 
\pi (-1)^m 
\int_{0}^{R_{\star}} 
Q^{1}_{(n \ell),(n' \ell')} 
\mathrm{d}r
\times 
\mathrm{cos} \, (o + o')  \nonumber \\
& + & 
\sum_{(n,\ell,m=0)} 
\sum_{(n',\ell'=\ell,m'=0)} 
2\pi 
\int_{0}^{R_{\star}} 
Q^{1}_{(n \ell),(n'\ell')} 
\mathrm{d}r 
\times 
\mathrm{cos} \, (o) 
\mathrm{cos} \, (o'). \nonumber \\  
\label{eq_A4}
\end{eqnarray}
It should be noted that the final term of the right-hand side in Equation~(\ref{eq_A4}) also contains the non-time-varying component 
but is not explicitly shown here just for simplicity. 

The second, third, and fourth 
terms in Equation~(\ref{eq19}) 
can be computed almost in the same manner, where Equations~(\ref{eq16}), (\ref{eq17}), (\ref{eq20}), and (\ref{eq21}) are used, 
resulting in almost the same expressions as Equation (\ref{eq_A4}) 
where $Q^{1}_{(n\ell),(n'\ell')}$ is replaced by 
$Q^{2}_{(n\ell),(n'\ell')}$, 
$Q^{3}_{(n\ell),(n'\ell')}$, 
and 
$Q^{4}_{(n\ell),(n'\ell')}$, 
respectively~(see Equations~(\ref{eq25_2}) to (\ref{eq25_4})). 
We can then obtain Equation~(\ref{eq24}) 
by introducing $Q_{(n \ell),(n' \ell')}$ (Equation~(\ref{eq25})). 

\section{Theoretical evaluation of the neutrino flux fluctuations caused by solar g modes in the case of CNO neutrinos} \label{App:CNO}
\begin{figure}[t]
\begin{center}
\includegraphics[scale=0.32]{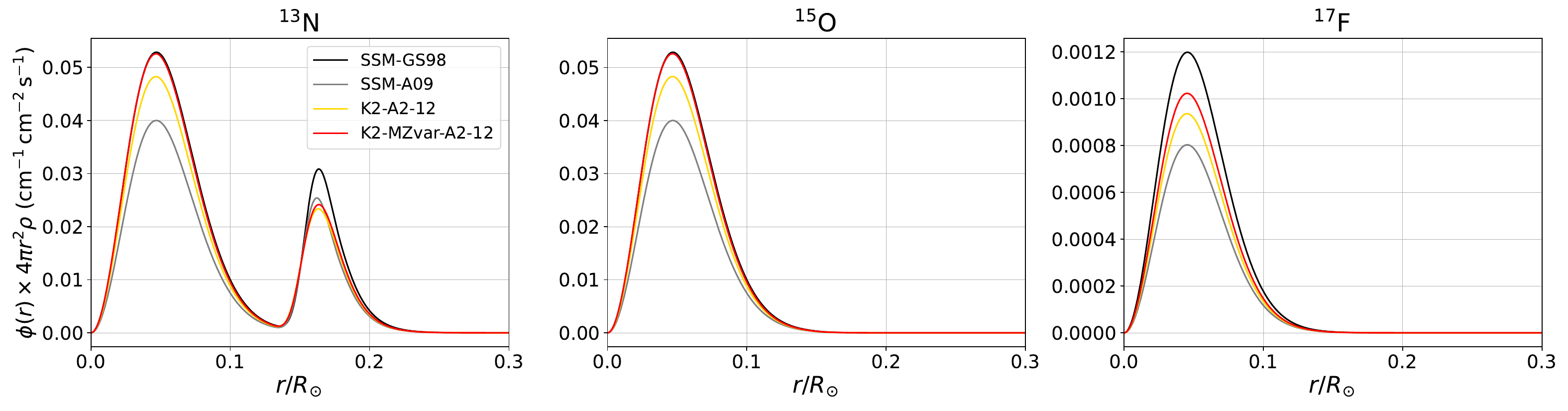}
\caption{\footnotesize Same as Figure \ref{fig:1} but for the CNO neutrinos, i.e., $^{13}\mathrm{N}$, $^{15}\mathrm{O}$, and $^{17}\mathrm{F}$ (from left to right).} 
\label{fig:CNO1}
\end{center}
\end{figure}

\begin{figure}[t]
\begin{center}
\includegraphics[scale=0.45]{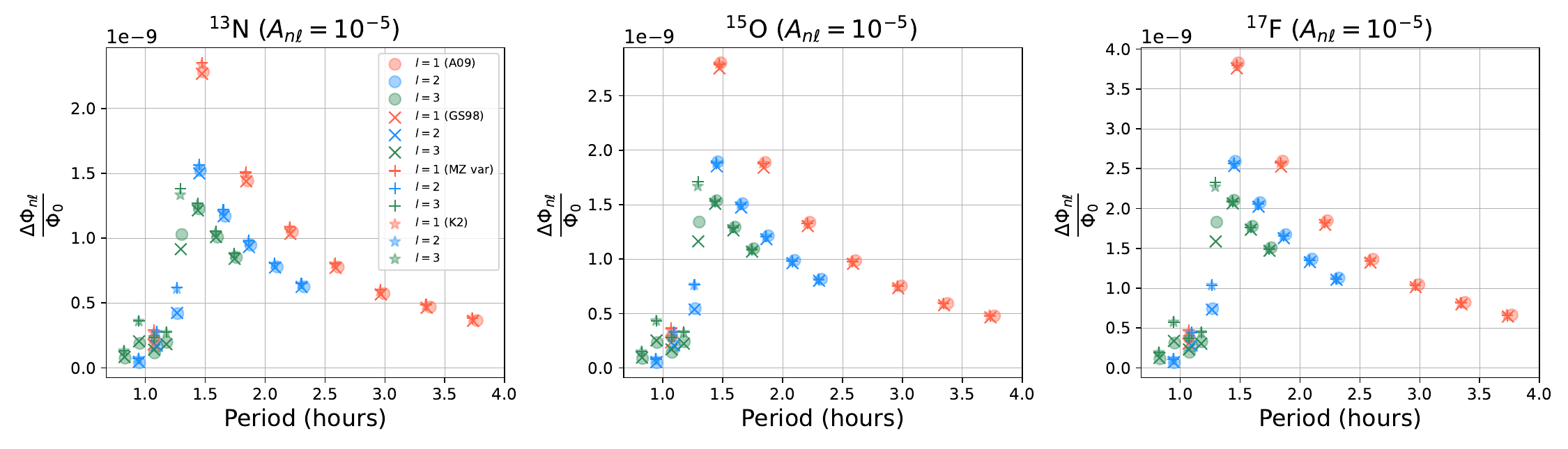}
\caption{\footnotesize Same as Figure \ref{fig:3} but for the CNO neutrinos, i.e., $^{13}\mathrm{N}$, $^{15}\mathrm{O}$, and $^{17}\mathrm{F}$ (from left to right).
}
\label{fig:CNO2}
\end{center}
\end{figure}

In this appendix, we show the results of numerical evaluation of the neutrino flux fluctuations caused by g modes in the case of the CNO neutrinos, i.e., $^{13} \mathrm{N}$, $^{15} \mathrm{O}$, and $^{17} \mathrm{F}$ whose fluxes per unit length are shown in Figure \ref{fig:CNO1}. 
We took the same approach for the numerical evaluation as described in Section~\ref{sec:3-3-1}, resulting in Figure~\ref{fig:CNO2}, where we actually see trends similar to those seen in Figure \ref{fig:3}. 

\begin{figure}[t]
\begin{center}
\includegraphics[scale=0.50]{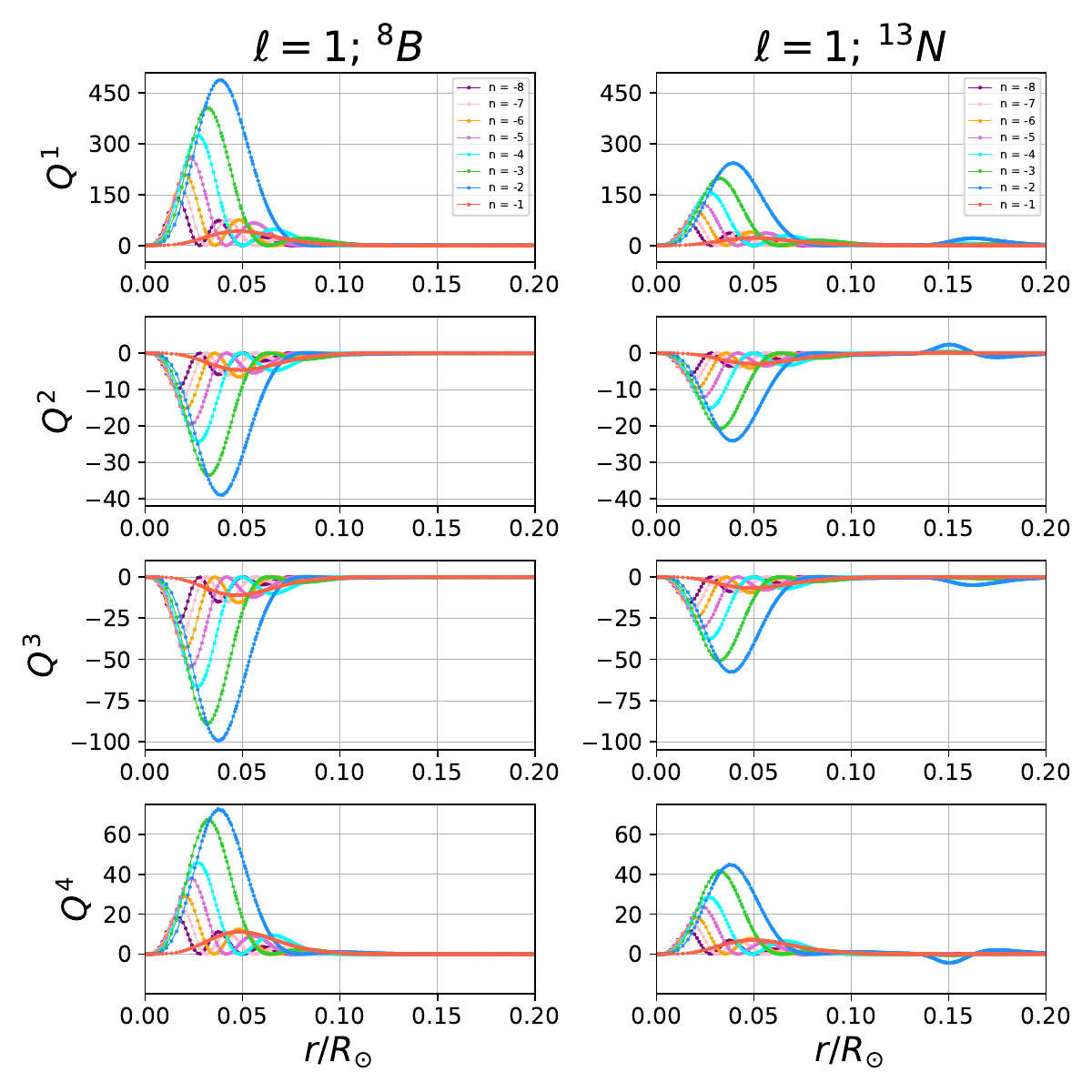}
\caption{\footnotesize The radial profiles of the functions $Q^{i}_{(n \ell),(n \ell)}(r)$ (defined in Equations (\ref{eq25_1}) to (\ref{eq25_4})) in the case of dipole ($\ell = 1$) modes with different radial orders ($n=-8$ to $-1$, represented by different colors). 
The left and right panels show the cases of $^{8}\mathrm{B}$ and $^{13}\mathrm{N}$ neutrinos, respectively. 
Although we have seen a clear difference in flux density profiles of $^{8}\mathrm{B}$ and $^{13}\mathrm{N}$ 
(see Figure \ref{fig:1}), 
the shapes of $Q^{i}_{(n \ell),(n \ell)}(r)$ in the case of $^{8}\mathrm{B}$ neutrino are fairly similar to those in the case of $^{13}\mathrm{N}$ neutrino, 
which can be attributed to the strong sensitivity of the squared eigenfunctions around the central region ($r/R_{\odot} < 0.1$). 
}
\label{fig:Ci}
\end{center}
\end{figure}

Nevertheless, it might be surprising that we see such similar trends in the period dependence of $\Delta \Phi_{\mathrm{g\text{-}mode}} / \Phi_0$ for various types of neutrinos even though the neutrino flux density profiles $\phi_0$ are different from one another (compare, e.g., the $^{8}\mathrm{B}$ in Figure \ref{fig:1} and $^{13}\mathrm{N}$ in Figure \ref{fig:CNO1}.) 
We can find the reason by taking a look at $Q^i_{(nl),(nl)}$ (Equations (\ref{eq25_1}) to (\ref{eq25_4})) computed for, e.g., $^{8}\mathrm{B}$ and $^{13} \mathrm{N}$ neutrinos (Figure \ref{fig:Ci}). 
It is clearly seen that, in the case of $^{13} \mathrm{N}$ neutrinos, the contributions of the outer bump in $\phi_0$ (see around $r/R_{\odot} \sim 0.17$ in Figure \ref{fig:1}) to $Q^{i}_{(n \ell),(n \ell)}$ are almost negligible (see around $r/R_{\odot} \sim 0.17$ in the right panels of Figure \ref{fig:Ci}). 
This suppression originates from the strong sensitivity of the squared eigenfunctions, such as $\delta T_{nl}(r)^2$, $\xi_{r,nl}^2$, etc., toward the center. 
As a result, regardless of kinds of neutrino, what contributes to $Q^{i}_{(n \ell),(n \ell)}$ is only the innermost bump in $\phi_0$, rendering $Q^{i}_{(n \ell),(n \ell)}$ profiles to be somewhat similar for different neutrinos.

\section{Asymptotic evaluation of the non-time-varying component in the neutrino flux fluctuation} \label{App:B}
In this appendix, we show that the non-time varying component ($\pi \int Q_{(n \ell),(n \ell)} \mathrm{d}r$) is positive for g-modes with $1 \le n \le 500$ and $1 \le \ell \le 500$.
For evaluating the integration, we need to compute the eigenfunctions ($\delta T$ and $\xi_r$). 
A straightforward way is to solve linear adiabatic oscillation numerically with, e.g., GYRE. 
However, since the numbers of meshes in the four solar models used in this study are around $\sim$ a few thousands, it might be possible that numerical computations would not work for high-order and high-degree modes. 
Such numerical inaccuracies may be avoided by somehow smoothing the structural variables, which nevertheless leads to inconsistency among the smoothed structural variables. 
Therefore, we take the asymptotic approach \citep[e.g.][]{Unno+1989}, where the g-mode property is mostly determined by the Brunt-V\"{a}is\"{a}l\"{a} frequency ($N^2 = -g (\frac{\mathrm{d} \, \mathrm{ln} \, \rho}{\mathrm{d}r} - \frac{1}{\Gamma_1} \frac{\mathrm{d} \, \mathrm{ln} \, p}{\mathrm{d}r})$, where $p$ and $\Gamma_1$ are the pressure and the first adiabatic exponent, respectively). 
We first by hand increase the number of the mesh points ($\sim 2 \times 10^{4}$) and smooth $N^2$ with which we compute the eigenfunctions basically following the formulations in \citet{Unno+1989}. 
We will describe the procedure in more detail in the following paragraphs. 
As we see little differences among the different solar models, we will show the case of our fiducial model (K2-MZvar-A2-12 in the main text). 
Please see \citet{Unno+1989} for more thorough discussions on asymptotic analysis of stellar nonradial oscillation. 

We start with determining the eigenfrequency and mode cavity for a mode labeled with a given $(n,\ell)$. 
This was done by using the asymptotic expression of the g-mode period: $P_{n\ell} = \frac{\Pi_0}{\sqrt{\ell(\ell+1)}} (n + 0.5 \ell + \alpha_g)$, where $\Pi_0$ is the inverse of the integration $\int_{r_1}^{r_2} N \mathrm{d} \, \mathrm{ln} \, r$, and the offset $\alpha_g$ is assumed to be $0.25$. 
Note that the integration in $\Pi_0$ is carried out for the mode cavity that is defined by a domain $[r_1,r_2]$ in which the squared radial wavenumber is positive, i.e.,: 
\begin{eqnarray}
k_r^2 = 
\frac{1}{\omega_{n \ell}^2 c^2} 
\biggl ( 
\omega_{n \ell}^2 - N^2
\biggr ) 
\biggl ( 
\omega_{n \ell}^2 - L_\ell^2
\biggr ) > 0 \label{eqB0}
\end{eqnarray} 
in the mode cavity. 
The sound speed and Lamb frequency (for a given $\ell$) are represented by $c$ and $L_\ell$, respectively. 
The eigenperiod and frequency are related to each other as: $\omega_{n \ell} = 2 \pi / P_{n \ell}$. 

Because the neutrino flux density $\phi(r)$ is concentrated around the central region ($r/R_\odot < 0.2$), we focus on the region around the inner turning point ($r \sim r_1$). 
A function $v$, which is proportional to $\xi_r$, may then be given as: 
\begin{eqnarray}
v 
= 
|k_r|^{-1/2} 
\biggl ( 
\bigl | 
\frac{3}{2} 
\int_{r_1}^{r} 
|k_r| 
\mathrm{d}r \bigl |
\biggr )^{1/6} 
A_i(\zeta), \label{eqB1}
\end{eqnarray}

\noindent where $A_i(\cdot)$ represents the Airy function and $\zeta$ is defined as: 
\begin{eqnarray}
\zeta = \mathrm{sgn}(k_r^2) 
\biggl ( 
\bigl | 
\frac{3}{2} 
\int_{r_1}^{r} 
|k_r| 
\mathrm{d}r \bigl |
\biggr )^{2/3}. \label{eqB2}
\end{eqnarray}

Once we compute $v$ for given $N^2$ and $(n,\ell)$, we can compute $\xi_r$ using the following approximate relation: 
\begin{eqnarray}
v = \frac{\omega_{n \ell}}{\sqrt{\ell (\ell+1)}}\rho^{\frac{1}{2}} r^2 \xi_r. \label{eqB3}
\end{eqnarray}

\noindent With a function $w$ that is defined as: 
\begin{eqnarray}
w = \rho^{-\frac{1}{2}} r 
(|N^2 - \omega_{n \ell}^2|)^{-\frac{1}{2}} p',  \label{eqB4}
\end{eqnarray}

\noindent and the definition of $\xi_h$ under the Cowling approximation: 

\begin{eqnarray}
\xi_h = \frac{p'}{\omega_{n \ell}^2 r \rho},  \label{eqB5}
\end{eqnarray}

\noindent we can subsequently obtain $p'$ and $\xi_h$. 

\begin{figure}[t]
\begin{center}
\includegraphics[scale=0.6]{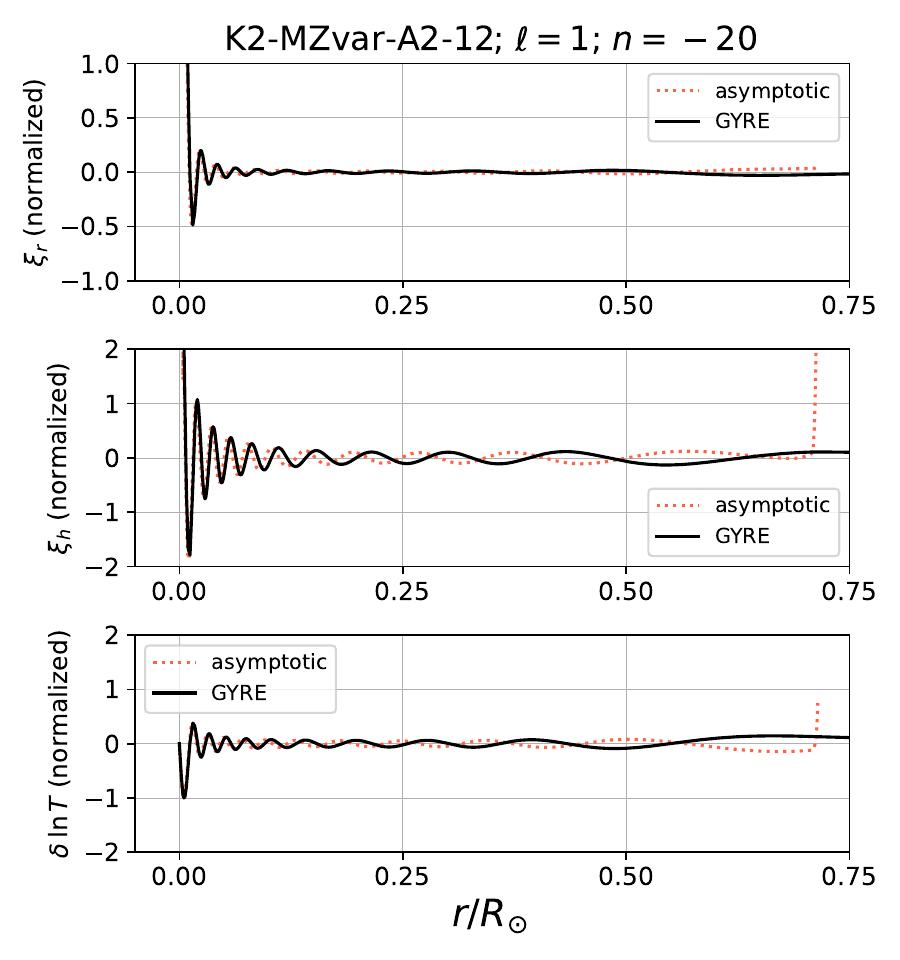}
\caption{\footnotesize Eigenfunctions $\xi_r$, $\xi_h$, and $\delta \, \mathrm{ln} \, T$ (from top to bottom) in the case of a g-mode with $(n,\ell) = (-20,1)$ obtained by numerically solving linear adiabatic oscillation via GYRE (black solid curves) 
and those obtained based on the asymptotic approach described in the main text (orange dotted curves). 
Note that the asymptotic eigenfunctions diverge around $r/R_{\odot} \sim 0.71$, which corresponds to the outer turning point 
at which the radial wavenumber $k_{r}^2$ changes sign from positive to negative. 
Since the neutrino flux density $\phi(r)$ is concentrated around the central region $r/R_{\odot} < 0.25$, the deviation does have negligible impacts on the numerical evaluation of the integration $\int Q_{(n \ell),(n \ell)} \mathrm{d}r$. }
\label{fig:B1}
\end{center}
\end{figure}

Finally, we calculate $\delta \rho$ via the adiabatic relation: 
\begin{eqnarray}
\delta \, \mathrm{ln} \, \rho 
= 
\frac{1}{\Gamma_1} 
\biggl ( 
p' + \xi_r \frac{\partial p}{\partial r}
\biggr ),  \label{eqB6}
\end{eqnarray}

\noindent which are then used to compute $\delta T$. 
We normalize the eigenfunctions thus obtained in the same manner as we described in Section~\ref{sec:3-2}. 
Just for comparison, we show in Figure~\ref{fig:B1} the eigenfunctions of a g-mode with $(n,\ell) = (-20,1)$ obtained by using GYRE and those obtained by the asymptotic approach explained above. 
We see more or less a nice agreement between them in the deep radiative region of the Sun. 

\begin{figure}[t]
\begin{center}
\includegraphics[scale=0.6]{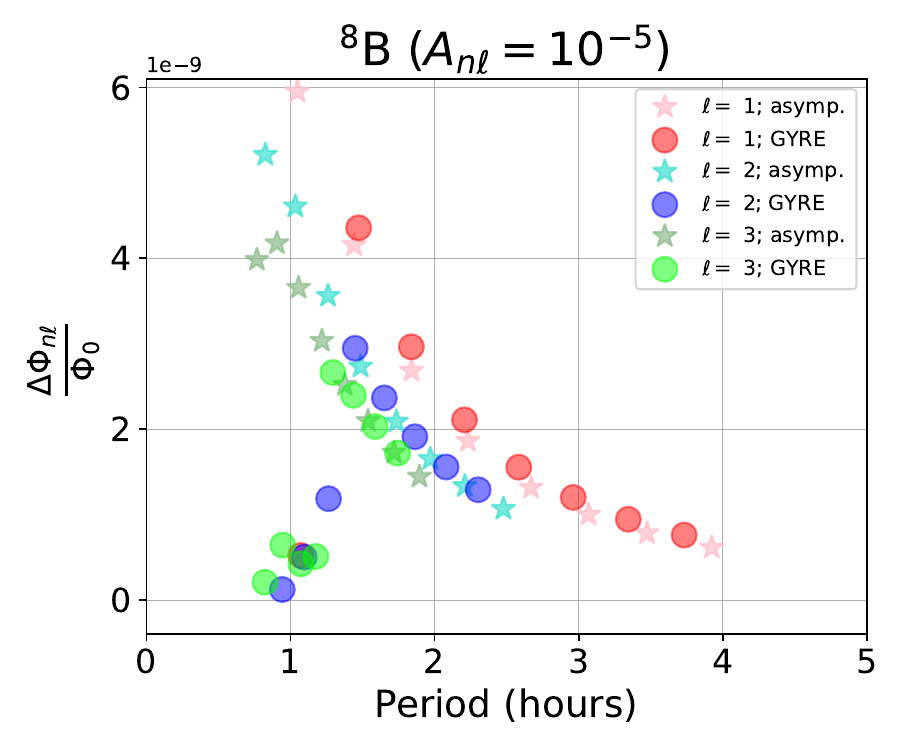}
\caption{\footnotesize Comparison of the non-time-varying component $\pi \int Q_{(n\ell),(n\ell)} \mathrm{d}r$ in the flux fluctuation of $^{8}\mathrm{B}$ neutrino 
that are evaluated with the eigenfunctions computed via GYRE (red, blue, and lime circles) 
and those evaluated with the asymptotic eigenfunctions (pink, turquoise, and moss green stars) 
as a function of the g-mode periods. 
The fiducial model (K2-MZvar-A2-12, see the main text) is used for the evaluations. 
Different colors indicate different spherical degrees. 
Except for the low-order ($n \le 4$) modes, 
we see a good agreement between the numerical and asymptotic evaluations. 
}  
\label{fig:B2}
\end{center}
\end{figure}

\begin{figure}[t]
\begin{center}
\includegraphics[scale=0.5]{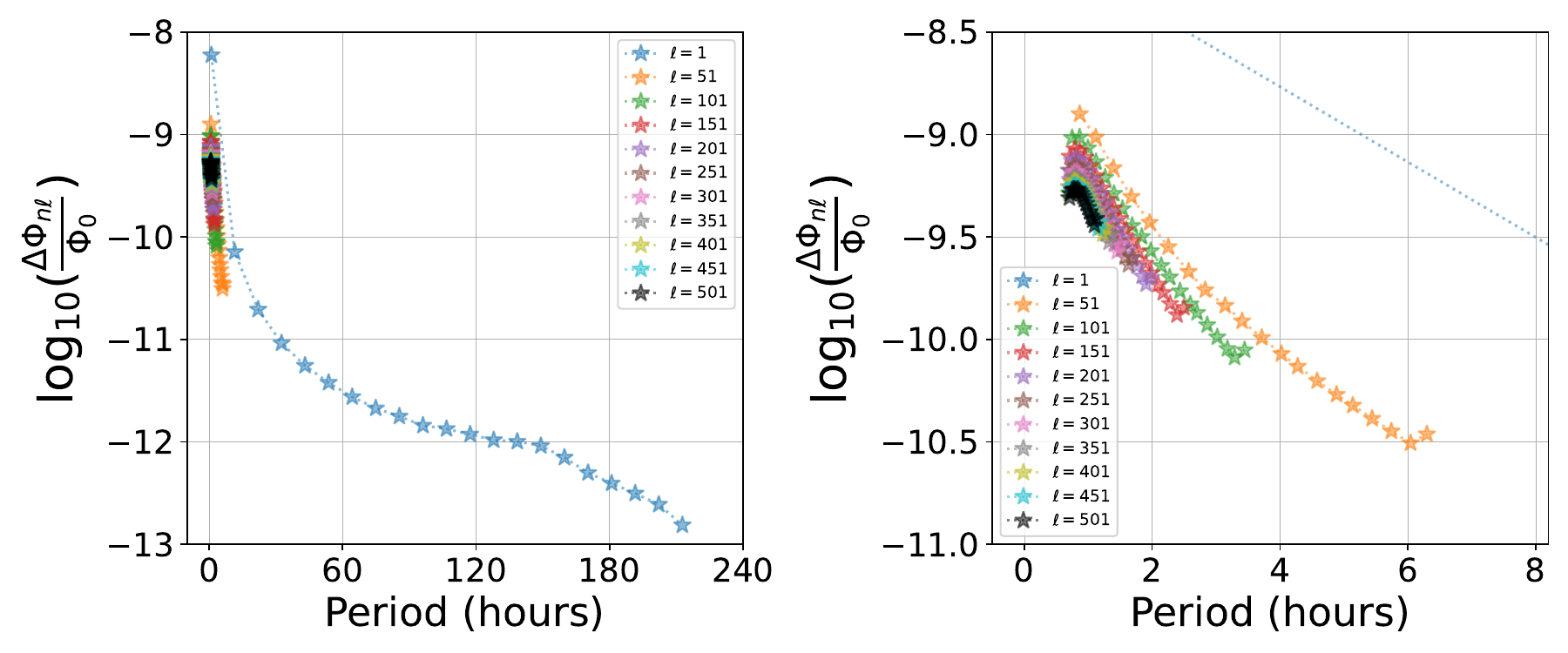}
\caption{\footnotesize Asymptotic evaluations of the non-time varying component $\Delta \Phi_{n \ell} / \Phi_0 = \pi \int Q_{(n \ell),(n \ell)} \mathrm{d}r$ 
as a function of the corresponding g-mode period (in hours) in the case of $^8\mathrm{B}$ neutrinos. 
The ranges of the radial order and the spherical degree are $1 \le n \le 501$ and $1 \le \ell \le 501$. 
For clarity, every twenty-five $n$ and 
every fifty $\ell$ are shown in the figure.  
The amplitude parameter $A_{n \ell}$ is assumed to be $10^{-5}$ for all the modes. 
Right panel is the expanded look into the shorter-period region ($0 < P_g<10$ hours) of the left panel. 
For all the g-modes evaluated here, the integration $\int Q_{(n \ell),(n \ell)} \mathrm{d}r$ is positive. 
Note that the vertical axis is represented in a logarithmic scale. 
}  
\label{fig:B3}
\end{center}
\end{figure}

\begin{figure}[t]
\begin{center}
\includegraphics[scale=0.5]{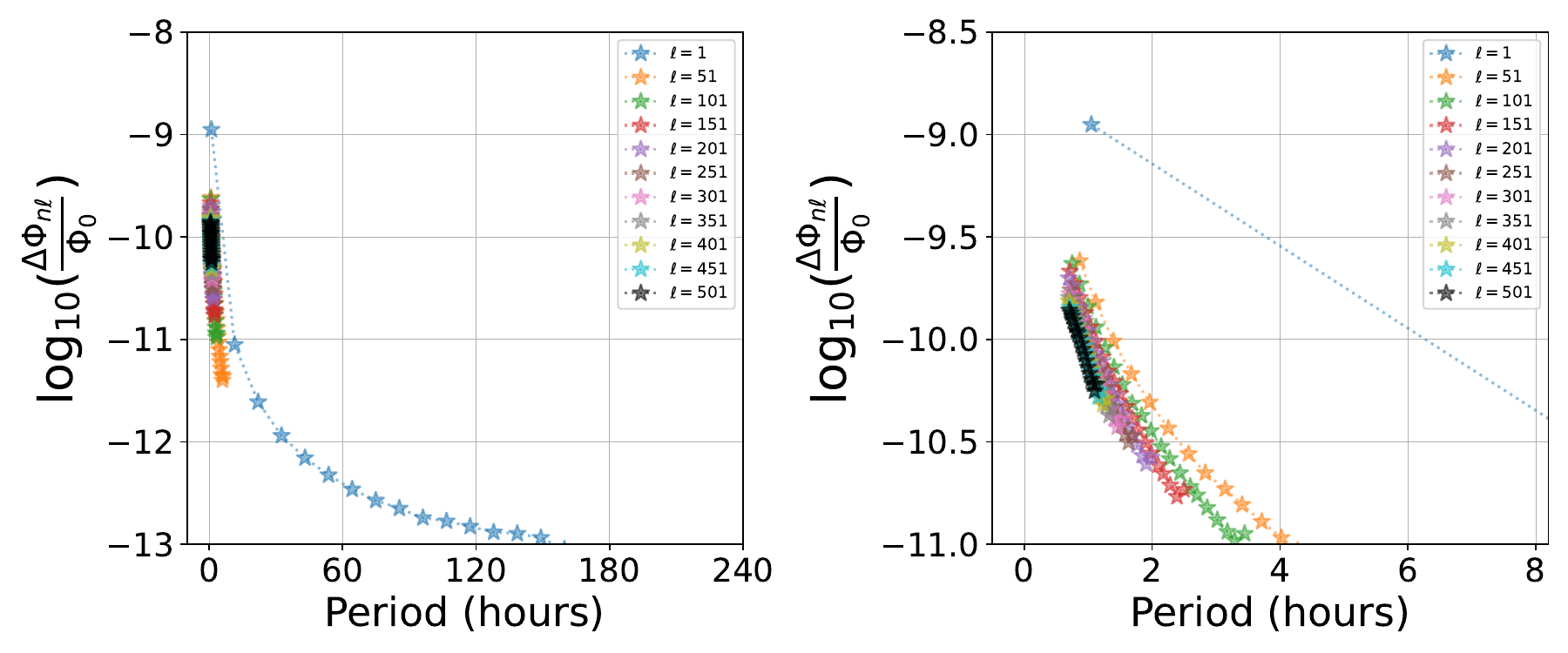}
\caption{\footnotesize Same as Figure \ref{fig:B3} in the case of $^7\mathrm{Be}$ neutrinos
} 
\label{fig:B4}
\end{center}
\end{figure}

We then numerically evaluate the integration $\int Q_{(n \ell),(n \ell)} \mathrm{d}r$ using the eigenfunctions thus computed. 
In Figure~\ref{fig:B2}, the evaluation results for low-degree and low-order g-modes in the case of $^{8}\mathrm{B}$ are plotted over the results shown in Section~\ref{sec:3-3-1} (Figure~\ref{fig:3}). 
It is clearly seen that the asymptotic results agree very well with the numerical ones especially for g-modes with radial orders higher than $4$. 
The deviation seen for lower-order modes ($n \le 4$) is expected because the asymptotic approximation is only applicable to high-order modes whose wavelengths are much shorter than the scale height of the background 
(the Brunt-V\"{a}is\"{a}l\"{a} frequency $N^2$ in this case.) 
We then present Figures~\ref{fig:B3} and \ref{fig:B4} in which the asymptotic evaluations for a wider range of mode indices~($1 \le n \le 500$ and $1 \le \ell \le 500$) are shown in the case of $^8\mathrm{B}$ and $^7\mathrm{Be}$ neutrinos, respectively. 
It is apparent that the integration $\int Q_{(n\ell),(n\ell)} \mathrm{d}r$ is positive 
for all the g-modes considered in this appendix, 
partly supporting our argument in Section~\ref{sec:3-3-3} that 
the net increase of the average neutrino flux~(caused by the second-order effects of g-mode oscillations) is simply proportional to the number of g-modes involved. 

\begin{figure}[t]
\begin{center}
\includegraphics[scale=0.3]{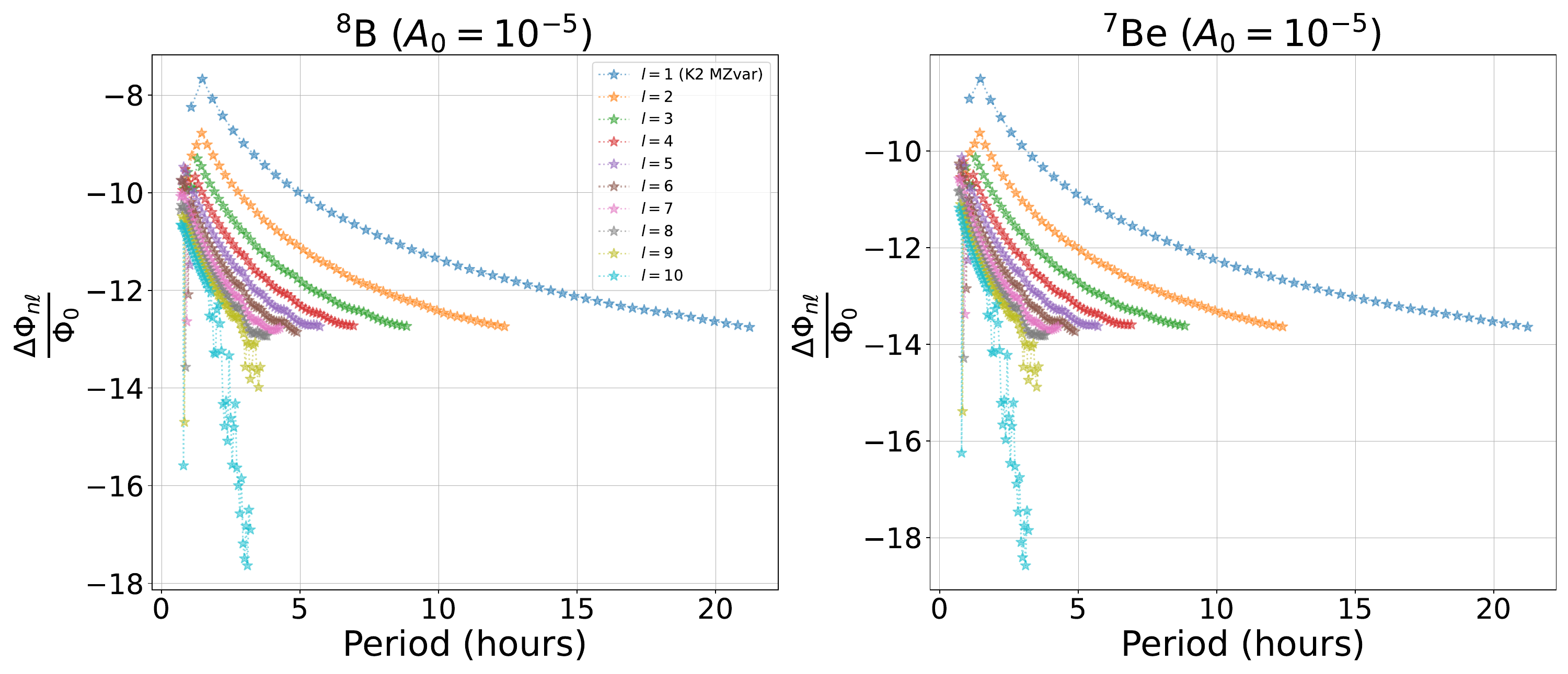}
\caption{\footnotesize \footnotesize Asymptotic evaluations of the non-time varying component $\Delta \Phi_{n \ell} / \Phi_0 = \pi \int Q_{(n \ell),(n \ell)} \mathrm{d}r$ 
as a function of the corresponding g-mode period (in hours) in the case of $^8\mathrm{B}$ (left) and $^7\mathrm{Be}$ (right) neutrinos. 
The ranges of the radial order and the spherical degree are $1 \le n \le 51$ and $1 \le \ell \le 10$. 
Note that we have assumed the dependence of $A_{n \ell}$ on the g-mode frequency and the spherical degree as: $A_{n \ell} = A_{0} \times (\nu_g / \nu_0)^{2.5} \times ( \ell (\ell + 1)/2 )^{-1}$ (see the text for more details on the variables), and the amplitude parameter $A_{0}$ is assumed to be $10^{-5}$. 
} 
\label{fig:BB}
\end{center}
\end{figure}

As discussed in the main text, the assumption that $A_{n \ell}$ is the same for all the g-modes may be somewhat strong, which possibly affects the total amount of the neutrino flux fluctuations (though the fact that the integration $\int Q_{(n\ell),(n\ell)} \mathrm{d}r$ is positive 
for all the g-modes is unchanged.) 
We therefore carried out numerical evaluation for the neutrino flux fluctuations using a different distribution for $A_{n \ell}$. 
Specifically, we have used the following ad hoc expression (see discussions in Section \ref{sec:neutrino3}): $A_{n \ell} = A_{0} \times (\nu_g / \nu_0)^{2.5} \times ( \ell (\ell + 1)/2 )^{-1}$, where $\nu_g$ and $\nu_0$ indicate g-mode frequencies and a certain g-mode frequency (here set to be $100 \, \mu \mathrm{Hz}$), respectively, and $A_{0}$ represents the relative amplitude (here assumed to be $10^{-5}$) for the corresponding g-mode. 
We show in Figure \ref{fig:BB} the results of numerical evaluation thus obtained. 
A narrow range of the radial order ($1 \le n \le 51$) and the spherical degree ($1 \le \ell \le 10$) is shown because the high-order (low-frequency) and the high-degree g-modes are predicted to have such small amplitudes that they are negligible. 
On the other hand, it should be noted that the low-order (high-frequency) and the low-degree g-modes have higher amplitudes compared with the case of uniform $A_{n \ell}$ for all the g-modes. 
Consequently, the total amount of the neutrino flux fluctuations may not be changed very much even if we consider different amplitude distributions. 


\section{How to estimate the total solar neutrino fluxes from the experimental data} \label{sec:nu-calc}

In this Appendix, we briefly summarize the calculation to estimate the amplitude of the long-term fluctuation of solar neutrinos observed in several neutrino experiments.

\subsection{Homestake experiment}

The Homestake experiment is the first generation of the solar neutrino detector, which was located at the Homestake gold mine in the United States~\citep{1998ApJ...496..505C}. It was operated for over $25$~years since $1970$, where this period spans the solar cycles~20, 21, and 22. The detector contains $615$~tons of $\mathrm{C_{2}Cl_{4}}$ as a target of solar neutrinos detection, and utilizes the absorption reaction, $\mathrm{^{37}Cl}+\nu_{e}\to\mathrm{^{37}Ar}+e^{-}$ with the reaction threshold is $0.814$~MeV. Because of the reaction threshold, flux intensities, and the energy dependence of the cross section, the reaction predominantly occurs by solar $\mathrm{^{8}B}$ neutrons~\citep{1996PhRvC..54..411B}. After about a month of exposure, the produced $\mathrm{^{37}Ar}$ is collected, and its radioactive decays are counted by a proportional counter to measure the solar neutrino flux.

We analyzed the full dataset of the capture rate taken in the Homestake experiment, where the interactions are predominated by solar $\mathrm{^{8}B}$ neutrinos. The capture rate is converted to the flux of solar $\mathrm{^{8}B}$ neutrino and then the amplitude of its fluctuation is evaluated. According to the fitting procedure with Equation~(\ref{eq:fit}), the amplitude~$A_{\mathrm{cycle}}^{\mathrm{Homestake}}$ is consistent with zero within its uncertainty and gives an upper limit of $A_{\mathrm{cycle}}^{\mathrm{Homestake}}<1.07\times 10^{6}~\mathrm{cm^{-2} \, s^{-1}}$ as summarized in Table~\ref{tb:fit}.

\subsection{SAGE, GALLEX, and GNO experiments}

Three experiments have used Gallium to confirm the solar neutrino problem: the SAGE experiment used $50$~tons of metallic gallium in Russia from 1990 to 2007~\citep{2009PhRvC..80a5807A}, the GALLEX experiment used $30.3$~tons in Italy from 1991 to 1997~\citep{GALLEX:1998kcz}, and the GNO experiment followed the complete of the GALLEX's operation from 1998 to 2003~\citep{2005PhLB..616..174G}. The absorption reaction of gallium, $\nu_{e}+\mathrm{^{71}Ga} \to \mathrm{^{71}Ge}+e^{-}$, is utilized to detect solar neutrinos. Because of its low energy threshold with $0.233$~MeV, the reaction is sensitive to all solar neutrinos, including $pp$, $\mathrm{^{7}Be}$, and $\mathrm{^{8}B}$ neutrinos. Considering the cross sections and fluxes of solar neutrinos, the interaction rates consist of about $55\%$ of $pp$ neutrinos, about $25\%$ of $\mathrm{^{7}Be}$ neutrinos, and about $10\%$ of $\mathrm{^{8}B}$ neutrinos~\citep{Bahcall:1987jc}.

We analyzed the capture rates taken in the three experiments above. The capture rate is converted to the flux of solar $\mathrm{^{7}Be}$ neutrinos because $pp$ neutrinos are not sensitive to the solar g-mode oscillation as described in Section~\ref{sec:3} and the fraction of the interaction rate due to  $\mathrm{^{8}B}$ neutrinos is small in the case of gallium capture reaction. According to the fitting procedure with Equation~(\ref{eq:fit}), the  amplitudes~$A_{\mathrm{cycle}}^{\mathrm{SAGE}}$ and $A_{\mathrm{cycle}}^{\mathrm{GALLEX/GNO}}$ are consistent with zero within their uncertainties and conservatively give their upper limits of $A_{\mathrm{cycle}}^{\mathrm{SAGE}}<0.83\times 10^{9}~\mathrm{cm^{-2} \, s^{-1}}$ and $A_{\mathrm{cycle}}^{\mathrm{GALLEX/GNO}}<1.11\times 10^{9}~\mathrm{cm^{-2} \, s^{-1}}$, respectively, as summarized in Table~\ref{tb:fit}.

\subsection{Super-Kamiokande experiment}

The Super-Kamiokande experiment, which is a water Cherenkov detector in Japan~\citep{2003NIMPA.501..418F}, started its operation in 1996. This detector observes solar $\mathrm{^{8}B}$ neutrinos via the neutrino-electron elastic scattering~\citep{Super-Kamiokande:1998qwk}. However, the distinction between solar neutrino signals and background events is difficult because the energy of such events overlaps with that of other background events, such as radioactive impurities in purified water~\citep{2020NIMPA.97764297N} and the spallation products by cosmic-ray muons~\citep{2016PhRvD..93a2004Z}. On the other hand, the Cherenkov light from the recoil electron after the elastic interaction can preserve the direction of the incoming neutrinos. Hence, the solar neutrino events are clearly observed over the background rate from the direction of the Sun.

The public data from the Super-Kamiokande collaboration is the list of the solar $\mathrm{^{8}B}$ neutrinos in the format of a 5-day sample between April 1996 and May 2018, including the pure water phases of SK-I, -II, -III, and -IV~\citep{2024PhRvL.132x1803A}. To extract the amplitude of solar $\mathrm{^{8}B}$ neutrinos flux from the Sun, we calculated the total solar $\mathrm{^{8}B}$ neutrino flux~$\phi_{\mathrm{Solar}}$, taking solar neutrino oscillation and cross section into account. The SK data contains $\nu_{\mu}$ and $\nu_{\tau}$ components in their solar neutrino sample because the elastic scattering channel is also sensitive to $\nu_{\mu}$ and $\nu_{\tau}$, whose cross section is about $6$~times smaller than that of $\nu_{e}$~\citep{1995PhRvD..51.6146B}. The interaction rate of $R_{\mathrm{SK}}$ is considered as 

\begin{equation}
R_{\mathrm{SK}} =  \phi_{\mathrm{Solar}}\left[ P_{ee} \sigma_{\nu_{e}}+P_{e\mu} \sigma_{\nu_{\mu}}+ P_{e\tau} \sigma_{\nu_{\tau}}\right]  \label{eq:interaction}
\end{equation}

\noindent where $P_{e \alpha}$ is the probability of neutrino oscillation, 
$\nu_{e} \to \nu_{\alpha}$~($\alpha=e, \mu, \tau$), and $\sigma_{\alpha}$ is the cross section for a neutrino with flavor $\alpha$. Here, we assume that the measured survival probability of electron neutrinos is $P_{ee}=0.333$ based on the latest solar neutrino analysis~\citep{2024PhRvD.109i2001A}, and the cross sections of $\sigma_{\mu}$ and $\sigma_{\tau}$ are the same. Considering the interaction rate, the estimated solar $\mathrm{^{8}B}$ electron neutrinos is expressed as $\phi_{\mathrm{SK}}\sim0.448 \, \phi_{\mathrm{Solar}}$. We conducted the fitting procedure with Equation~(\ref{eq:fit}) using $\phi_{\mathrm{Solar}}$ data and found that amplitude~$A_{\mathrm{cycle}}^{\mathrm{SK}}$ is consistent with zero within its uncertainty and gives an upper limit of $A_{\mathrm{cycle}}^{\mathrm{SK}}<0.0675\times 10^{6}~\mathrm{cm^{-2} \, s^{-1}}$ as summarized in Table~\ref{tb:fit}.

\subsection{Borexino experiment}

The Borexino experiment, which is the organic liquid scintillator detector in Italy~\citep{Borexino:2008gab}, started its operation on 2007 and completed on 2021. The Borexino detector has observed many kinds of solar neutrinos~\citep{2011PhRvL.107n1302B, Borexino:2011ufb, Borexino:2017uhp, BOREXINO:2018ohr, 2020Natur.587..577B} and contributed to solve the inner composition problem~\citep{2009ApJ...705L.123S, Asplund+09, 2010Ap&SS.328...13S, 2019FrASS...6...42B, Orebi-Gann+21, 2022A&A...661A.140M}. The Borexino detector has reduced its internal background with several techniques, and the count rate of solar $\mathrm{^{7}Be}$ neutrino interactions clearly demonstrates the annual modulation after subtracting its estimated background rate.

The Borexino collaboration provides its solar $\mathrm{^{7}Be}$ neutrinos data from 2012 to 2021, which includes Phase-II and Phase-III, on their web page~\citep{2023APh...14502778A}. Note that the Phase-I data is not provided due to the high background rate originating from the internal radioactive impurities, such as $\mathrm{^{210}Po}$, $\mathrm{^{210}Bi}$, and $\mathrm{^{85}Kr}$~\citep{Borexino:2013zhu}. 

In order to evaluate the amplitude of solar $\mathrm{^{7}Be}$ neutrinos, we analyzed the publicly available data provided by the Borexino collaboration, which lists the count rate after subtracting the background events. Then, we fitted the two sine curves, defined in Equation~(\ref{eq:fit}). Since the interaction to detect solar $\mathrm{^{7}Be}$ neutrinos in the Borexino detector is the elastic scattering, the interaction rate is calculated based on the same manner as in the SK described in Equation~(\ref{eq:interaction}) with the survival probability of electron neutrino is $P_{ee}=0.53$ assuming the latest oscillation parameters. According to the fitting procedure with Equation~(\ref{eq:fit}), the amplitude~$A_{\mathrm{cycle}}^{\mathrm{Borexino}}$ is consistent with zero within its uncertainty and gives an upper limit of $A_{\mathrm{cycle}}^{\mathrm{Borexino}}<0.07\times 10^{9}~\mathrm{cm^{-2} \, s^{-1}}$ as summarized in Table~\ref{tb:fit}.

\subsection{SNO experiment}

The SNO experiment, which is the heavy water Cherenkov detector in Canada~\citep{2000NIMPA.449..172B}, started in 1999 and completed its operation in 2006. The SNO experiment uses deuterons~($\mathrm{^{2}H}$) as a target for the neutrino interactions~\citep{1985PhRvL..55.1534C}, and solar neutrinos undergo three different kinds of interactions, such as elastic scattering, charged current, and neutral current interactions. As a consequence, the SNO detector can measure not only the pure electron neutrino flux but also the total solar neutrino flux including all neutrino flavors~\citep{2007PhRvC..75d5502A}.

The SNO collaboration provides the list of the detection time of observed events in the Phase-I~(pure $\mathrm{D_{2}O}$, $572.2$~days) data from November 1999 to May 2001 and Phase-II~(Salt, $312.9$~days) data from July 2001 to August 2003~\citep{2005PhRvD..72e2010A, 2010ApJ...710..540A}. We should note that this dataset contains about five years and this duration is not long enough to evaluate the amplitude of the 11-year periodic change discussed in this article. For this reason, we did not include the SNO result in this analysis and we hope to request the SNO collaboration to share the full dataset, including the phase-III.

\end{document}